\documentclass{aa}  

\usepackage[position=top]{subfig}
\usepackage{graphicx}
\usepackage{csquotes}
\usepackage{txfonts}
\usepackage{svg}
\usepackage[T1]{fontenc}
\usepackage{xcolor}
\usepackage[colorlinks=true,linkcolor=blue,allcolors=blue]{hyperref}
\usepackage[capitalise]{cleveref}
\definecolor{blackberry}{HTML}{8D1D75}
\definecolor{lightblue}{rgb}{0.1,0.5,0.89}

\usepackage{float}

\begin{document}

\title{Go with the Flow: The Self-Similar and Non-Linear Behaviour of Large-Scale In- and Outflows and the Impact of Accretion Shocks from Galaxies to Galaxy Clusters}
\titlerunning{In- and Outflows from Galaxies to Galaxy Clusters}

   \author{Benjamin A. Seidel\inst{1}\thanks{E-mail: bseidel@usm.uni-muenchen.de}
           \and
           Rhea-Silvia Remus\inst{1}
           \and
           Lucas M. Valenzuela\inst{1}
           \and
           Lucas C. Kimmig\inst{1}
           \and
           Klaus Dolag\inst{1}\fnmsep\inst{2}
          }

   \institute{Universitäts-Sternwarte, Fakultät für Physik, Ludwig-Maximilians Universität München, Scheinerstr.1, 81679 München, Germany
   \and
   Max-Planck-Institute for Astrophysics, Karl-Schwarzschild-Str.\ 1, 85748 Garching, Germany
             }

   \date{}

 
  \abstract
   {From the scale-free nature of gravity, the structure in the Universe is expected to be self-similar on large scales. However, this self-similarity will eventually break down due to small-scale gas physics such as star formation, AGN and stellar feedback as well as non-linear effects gaining importance relative to linear structure formation. In this work we investigate the large-scale matter flows that connect collapsed structures to their cosmic environments specifically for their agreement with self-similarity in various properties. For this purpose we use the full power of the hydrodynamical cosmological simulation suite Magneticum Pathfinder to precisely calculate the instantaneous in- and outflow rates for structures on a large range of masses and redshifts. We find a striking self-similarity across the whole mass range and through time that only breaks down in the outflowing regime due to the different outflow driving mechanisms for galaxies vs. galaxy clusters. Geometrical analysis of the patterns of in vs. outflow demonstrate how the inflows organize into anisotropic filaments driven by the tidal distortions of the environment, while the outflows are fairly isotropic due to their thermal nature. This also manifests in the differences in the thermal and chemical properties of the gas in the in- and outflowing component: While the inflowing gas is pristine and colder, encountering the accretion shock surfaces and entering the influence region of AGN and stellar feedback heats the gas up into a diffuse, metal enriched and hot atmosphere. Overall the differences between outflowing and infalling gas are enhanced at the galaxy cluster scale compared to the galaxy scale due to the strong accretion shocks that reach out to large radii for these objects. An individual study of the gas motions in the outskirts of one of the most massive clusters in the simulations demonstrates these results to greater detail: Gas found in the outer ($r>1.2r_\mathrm{vir}$) hot atmosphere at $z=0$ falls in and is completely enriched early in the assembly process before being shock heated and expanding.}

   \maketitle
%
\section{Introduction}

A detailed understanding of the multi-scalar process of structure formation remains a key objective for modern cosmology and astrophysics. Since the notion of the inter-connectivity of the large scale cosmic web was first conceived by Zeldovich in his seminal paper \citep{zeldovich1970a}, it has become ever more clear that the large-scale assembly of matter into a complex web of filaments, sheets and nodes has consequences that proliferate well into the regime of non-linear structure formation and evolution in the galaxy regime \citep[e.g.][]{papovich2018a,castignani2022}. Especially the filamentary and smooth gas flows in close vicinity to galaxy clusters and galactic haloes are governed by a complex combination of the gravitational environment they arise in and the internal gas physics that predominantly drive the outflows.  A comprehensive overview of the complex physics in the outskirts of galaxy clusters is presented by \citet{walker2019}. 

On the galaxy scale, the make up of the large-scale environment is widely assumed to be a crucial factor in its star formation history, possible quenching and the galaxy's dynamical evolution \citep{galarraga-espinosa2023}. If the gravitational Universe is truly self-similar and the morphological differences at small scales arise purely from the increased importance of baryonic processes at smaller scales, the configuration of the gas and dark matter flow fields around galaxy clusters and galaxies must bear some residual resemblance in global properties, albeit scaled by the different gravitational energies involved.
The outskirts of (quasi-) virialized structures, where infalling gas and smaller structures encounter the thermalized medium of the already assembled larger structure for the first time are a critical test bed for our understanding of how these structures form and accrete mass over time. Both the hot components of the intra-cluster medium (ICM) in the outer regions of galaxy clusters and the multi-phase circum-galactic medium (CGM) have been the focus of large observational \citep[e.g.][]{gatkine2019,gatuzz2023} and theoretical \citep[e.g.][]{churazov2023,bogdan2023} efforts in the recent past. 
A better understanding of the ICM and its complex non-thermal physics, such as clumping, bulk flows and magnetic fields, is critical for improving mass estimates obtained from X-ray observations for cosmology. These non-thermal processes can introduce significant biases when assuming hydrostatic equilibrium to calculate masses. This can significantly impact the constraining power of cluster masses on cosmological parameters \citep{salvati2019,lebeau2024c}. 
 \begin{figure*}
\centering
\subfloat{
  \includegraphics[width=0.45\textwidth]{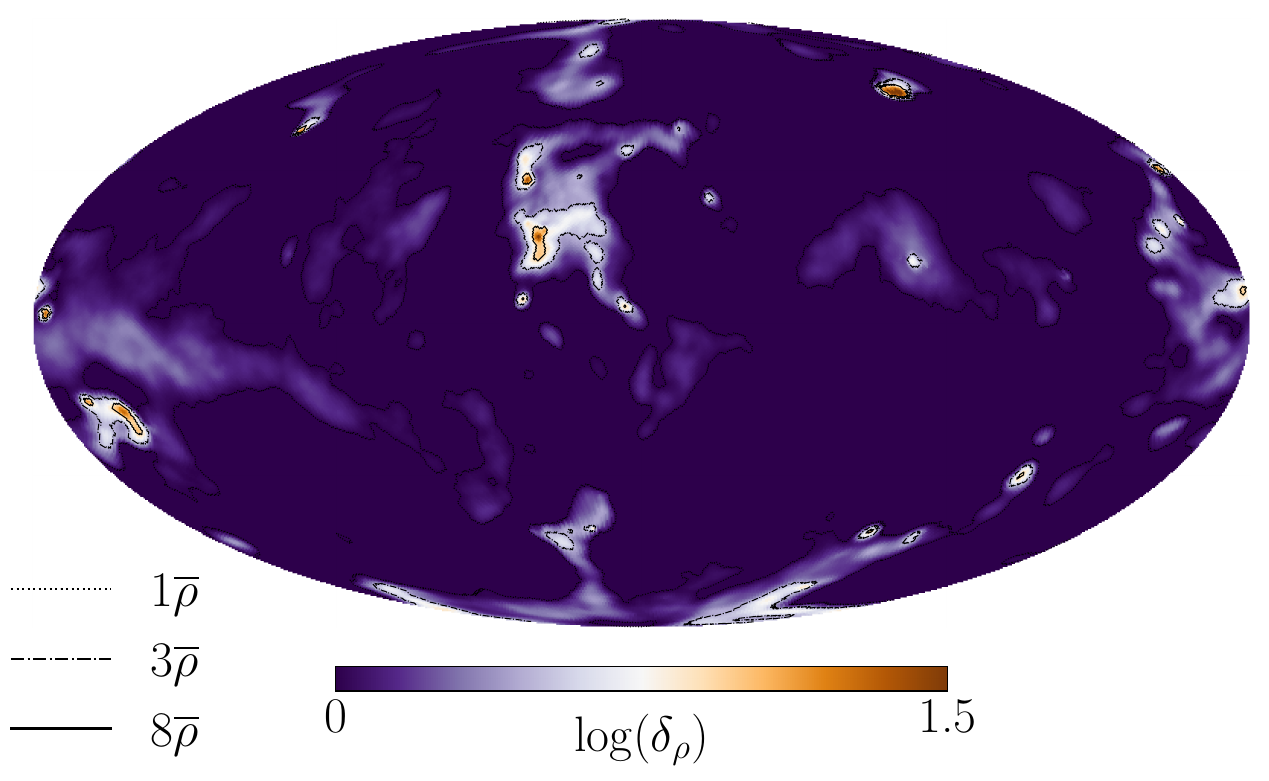}
}
\subfloat{
  \includegraphics[width=0.45\textwidth]{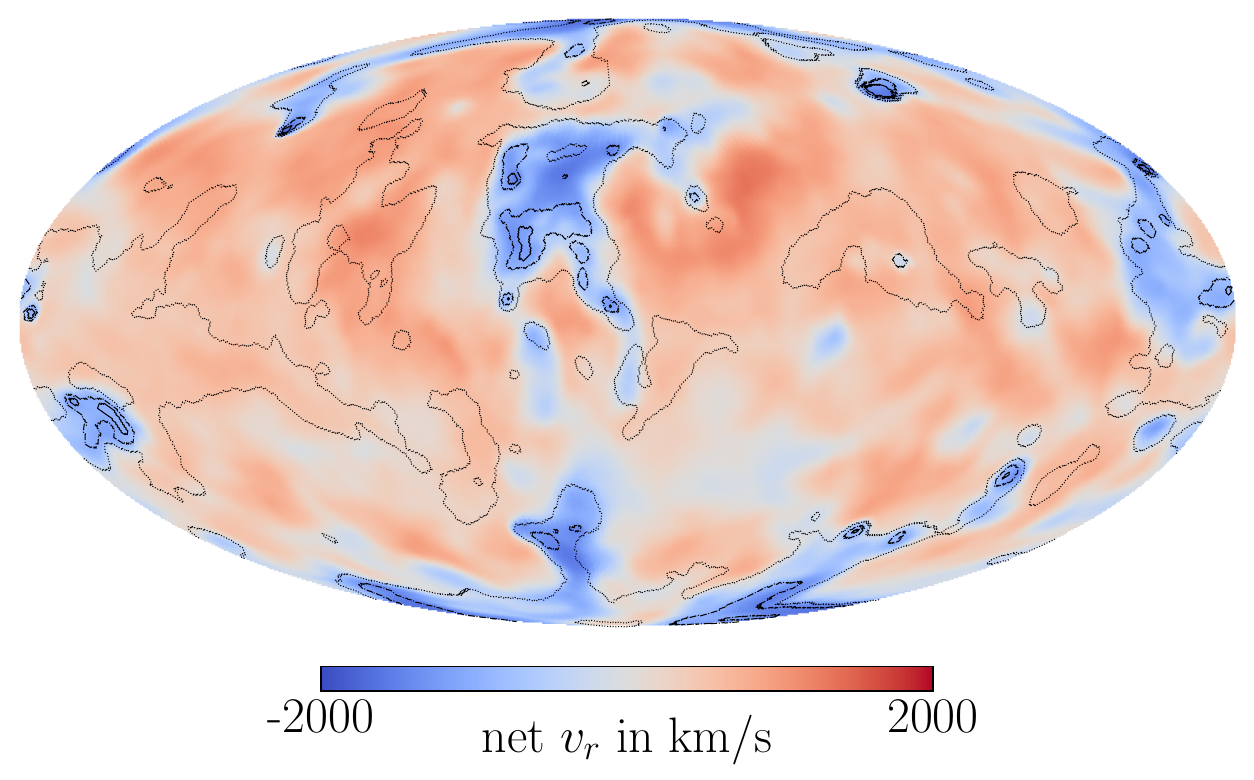}
}
\hspace{0mm}
\subfloat{
  \includegraphics[width=0.45\textwidth]{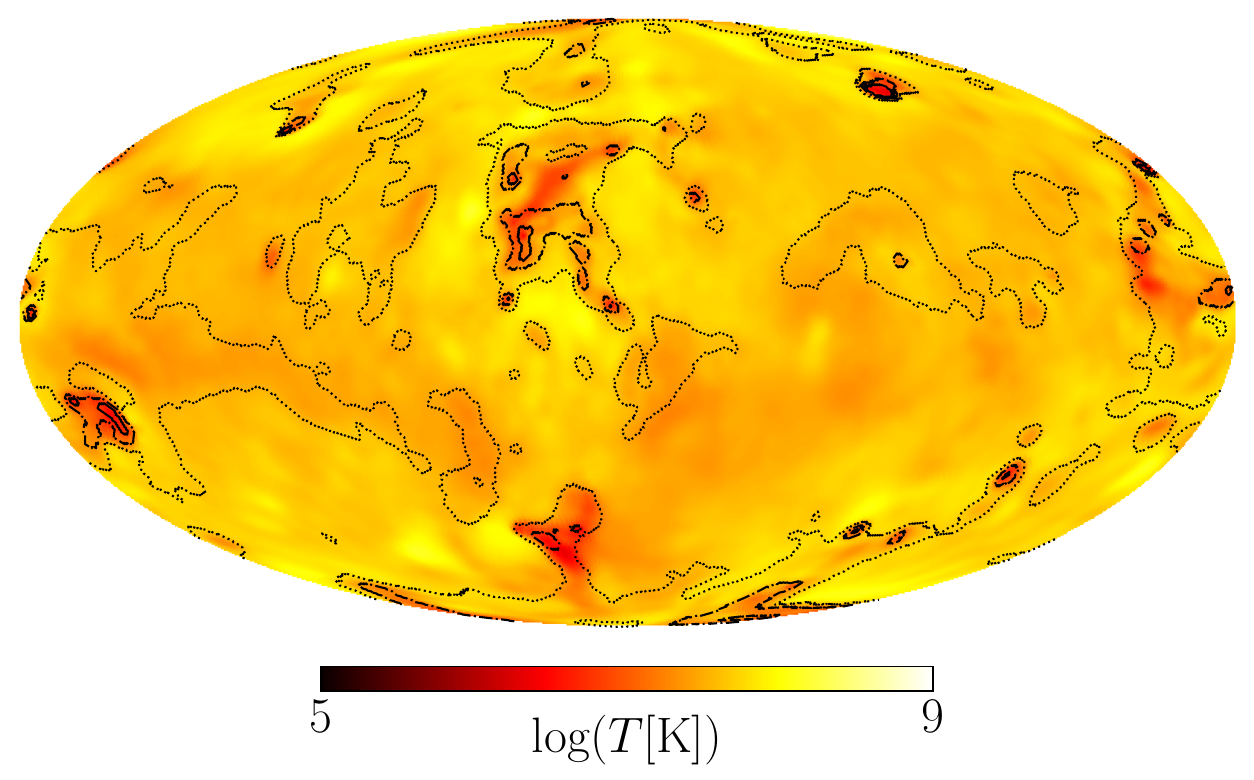}
}
\subfloat{
  \includegraphics[width=0.45\textwidth]{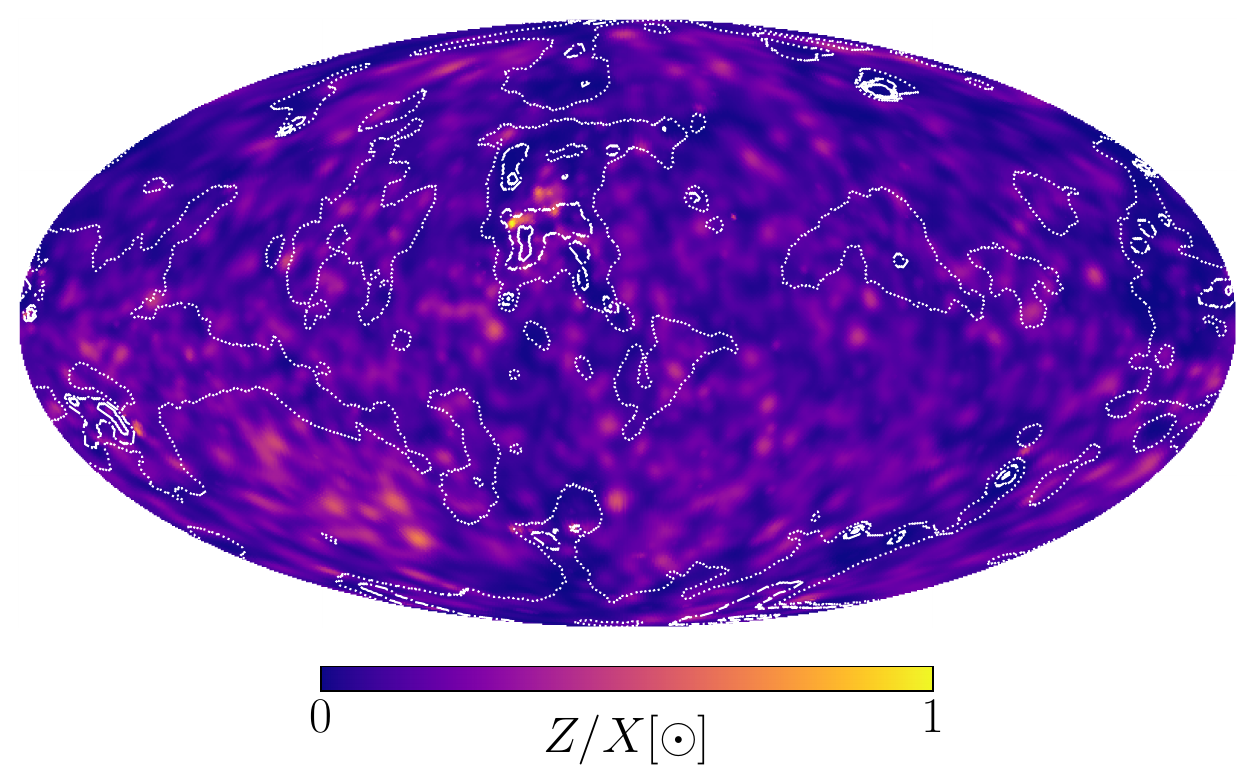}
}

\caption{The four fields (density, radial velocity, temperature and metallicity) interpolated on the spherical shells just outside the virial surface ($r=1.2r_{\mathrm{vir}}$) shown for an exemplary cluster from the highest mass bin ($10^{15}M_{\odot}$). The contours indicate the overdensity at the mean density of the shell $\bar{\rho}$, $3\bar{\rho}$ and the 
$8\bar{\rho}$ level respectively.}
\label{molly}
\end{figure*}
In the standard self-similar assembly picture the radius at which the in-falling gas is shock-heated by the already virialized atmosphere (accretion shock radius, $r_\mathrm{sh}$), and the apoapsis of the first-infall orbit of dark matter (splash-back radius, $r_\mathrm{sp}$) are generally expected to coincide \citep{shi2016,dekel2006}. However, recent numerical simulations have demonstrated that encounters of merger driven shocks with the accretion shock and the resulting merger-accelerated shock (MA-shock) can drive baryonic shocks out to multiple virial radii, far exceeding $r_\mathrm{sp}$ \citep{zhang2020} and create complex non-spherical baryonic boundaries with gas properties rapidly changing across these surfaces.   

 The CGM on the other hand is a key part in understanding how galaxies regulate their star formation in a complex interplay of large-scale gas accretion, cooling flows from the cold phase CGM and thermal feedback from star formation and active galactic nuclei (AGN) \citep{yu2021,stuber2021,hogarth2021}.  The balance between pristine inflowing gas from the cosmic web feeding the star formation in galaxies, and outflowing gas heated and enriched by feedback caused by the star forming activity and the central AGN is still not fully understood. Nevertheless, it is most crucial for understanding the evolution of galaxies and their properties through cosmic time. 
 
 A holistic  theoretical understanding of structure formation and how collapsed structures are connected to their environment is in general crucial in the exploration and interpretation of the increasingly precise X-ray and SZ observations of the present and near future. Additionally it can aid in interpreting the highly resolved observations of inflows, outflows and feedback bubbles now resolved in many galaxy observations \citep{watkins2023}. From the purely gravitational perspective, structure formation is assumed to be self-similar, that is structures assemble the same way at all scales, owing to the scale-invariant nature of gravity. It is then precisely the \textit{deviation} from self-similarity that encodes interesting information on the complex baryonic processes that set the evolution of massive and less massive haloes apart. Balancing the reproduction of observed CGM and ICM properties simultaneously poses a formidable challenge to numerical simulations, nevertheless it is integral to their predictive power.  
This work aims to tackle the mass assembly of organised structures in the Universe from a multi-scalar perspective and provide insights on the interaction of physical processes separated by several orders of magnitude. To this end we investigate the mass flows in the outskirts of objects from galaxy (group) to cluster masses and analyse their gravitational and thermodynamical properties in the framework of a state-of-the-art hydrodynamical cosmological simulation.      

This paper is structured as follows: In \cref{Sec:Methods} we introduce the numerical and theoretical methods we use to compute the flow fields in the Magneticum simulations and introduce the sample of haloes we apply these methods to. In \cref{Sec:Flows} we discuss the accretion flows and outflows across the full mass range as well as their geometric properties. We then connect these flows to the entropic structure of the outskirt gas and specifically the shocks in \cref{Sec:Shocks}. Additionally we follow accreting gas particles in the outskirts of a massive cluster as they encounter those shocks and study their thermal and chemical properties. Finally, in \cref{Sec:Phase} we discuss the phase properties of infalling and outflowing gas across the whole mass range in light of the findings of the preceding case study. 

\section{Methods} \label{Sec:Methods}
 
\subsection{Simulations}
We use the Magneticum Pathfinder suite of simulations\footnote{www.magneticum.org} which features a large selection of simulation volumes and resolutions (Dolag et al. 2025, to be submitted). All Magneticum Pathfinder simulations were performed with an updated version of the parallel cosmological Tree-PM code {\sc GADGET-2} \citep{springel2005a}, namely {\sc P-GADGET3}. To simulate the hydrodynamics, a modified smoothed-particle hydrodynamics (SPH) scheme was employed, with modifications such as improved thermal conduction \citep{dolag2004} and a low-viscosity formulation of SPH \citep{dolag2005b, beck2016a} specifically to improve the modelling of ICM physics. Subgrid physics include metal dependent cooling, star formation and detailed treatment of stellar evolution \citep{tornatore2004,tornatore2007} with supernova driven winds \citep{springel2003a}, black hole growth and feedback from AGN \citep{springel2005b,fabjan2010,hirschmann2014}. 

Magneticum Pathfinder offers a number of different simulation runs, varying in size and resolution (and implemented physics). For the purpose of this work we used the largest high resolution box ($2\times2880^3$ particles), {\it Box2b/hr}. Past studies of galaxy cluster physics in this box include (but are not limited to): Studies of galaxy cluster substructures \citep{kimmig2023}, studies of older, low-surface brightness galaxy clusters \citep{ragagnin2022}, galaxy populations and star formation quenching due to cluster environments \citep{lotz2019,lotz2021a} and predictions for soft X-ray excess emissions from the outskirts of galaxy clusters and the filamentary warm-hot intergalactic medium (WHIM) \citep{churazov2023}. The evolution of protoclusters from z=4 to the present day was investigated by \citep{remus2023}. Additionally a smaller box with higher resolution, {\it Box4/uhr}, was used to explore the lower end of the mass scale. This box has been primarily employed in galaxy studies such as global angular momentum properties \citep{teklu2015} and accreted \citep{remus2022} and baryonic \citep{harris2020} mass fractions of galaxies. Most relevantly for this work, it has recently been used to study the connection between star formation and gas accretion from the cosmic web (Fortuné et al. submitted) The cosmological parameters of the runs used are: $h=0.7$, $\Omega_m=0.272$, $\Omega_{\Lambda}=0.728$, $\sigma_8=0.809$, $n=0.963$ consistent with the WMAP-7 cosmology \citep{komatsu2011a}. 

\subsection{Identifying the systems}

{\sc GADGET3} includes an on-the-fly structure finding toolkit \textsc{SubFind} \citep{springel2001a} with modifications to the substructure identification \citep{dolag2009a}, which creates detailed Friends-of-Friends (FOF)-based catalogues of clusters and sub-cluster child structures (that is galaxies). We sampled a Catalogue of $29\times11 = 319$ randomly selected haloes in 11 logarithmic mass bins  from $10^{12}-10^{15}M_{\odot}$. The logarithmic width of each bin was was selected to ensure that at least 29 objects existed in each mass bin \footnote{the number 29 was motivated by the 29 most massive clusters from the high-resolution box ({\it Box2b/hr}) as these are the haloes with mass $M_\mathrm{vir}>10^{15}M_\odot$, see \citet{kimmig2023} for further details}. We connect {\it Box2b/hr} and {\it Box4/uhr} at the $10^{13}M\odot$ mass bin, where {\it Box2b/hr} still has enough resolution to resolve the haloes sufficiently and {\it Box4/uhr} has enough haloes to build a mass bin with a sufficiently small mass spread. Dolag et al. 2025 show the resolutions to converge in the most relevant scaling relations for galaxies and clusters indicating that this mixing of resolutions is justified. 

When selecting haloes at different timesteps, we did not trace individual haloes back in time but sampled new bins at each redshift, keeping the associated bin mass and not its constituents constant. The selected haloes are exclusively selected from the field, that is they are not bound to another, larger halo at some scale. This selection specifically excludes galaxies and groups evolving in a cluster environment, which are not the focus of this paper.   

\subsection{Reconstructing the peculiar mass flow from simulation data}
The flow through a surface given a continuous velocity field $\vec{v(x)}$ can be generally computed by either integrating its divergence over the enclosed volume or via the Gaussian theorem by
\begin{equation}
\label{1}
\int_V \vec{\nabla} \cdot \vec{F} \,dv = \oint_{\partial V} \vec{F} \cdot \hat{n} \,da
\end{equation}. From a numerical perspective the latter method is the more sensible (integrating over a volume, scaling with $r^3$, vs. a surface, with $r^2$ scaling). The largest haloes in the sample reach several Mpc in virial radius or in other words scales where the expansion of the background cosmology becomes relevant, therefore we use \textit{peculiar} velocities throughout this work.
In SPH the gas discretisation is based on representative point particles that are smoothed over a volume of space with a density dependent kernel. To accurately model the gas flow given a set of gas \enquote{particles} it is therefore necessary to take into account the geometry and smoothing lengths of the kernels with the advantage that SPH will provide a continuous gas velocity field without artificial clumpiness. To obtain a similarly smooth velocity field for the dark matter component, equivalent smoothing lengths are computed in post-processing for this species by calculating a weighted mean density at each particle position using a simple neighbour tree.

\subsection{Geometry}
Because dark matter haloes can to first order be considered as approximately spherically symmetric (deviations from this assumption wrt. to the matter configuration in the outskirts are discussed in section \ref{fgeo}) for the purpose of this investigation, we compute the flow through spherical shells. We set up the computation of the inflow and outflow rates to be at spherical surfaces of $r\in [1.2,1.4,.....,5.0] \times r_{\mathrm{vir}}$. The peculiar particle velocities were corrected by subtracting the centre of mass motion of the main halo the shells are centered on.
\subsection{Flowmaps}
We use the program {\sc SMAC} \citep{dolag2005a} to generate the flow maps. Our custom version of this post-processing software -- adding a slice based computation of the quantities for flow computation as opposed to the standard radial binning method - uses a healpix \citep{gorski2005}  based tessellation of the surface to compute the SPH based gas density at every pixel of the spherical shell. This density is calculated based on the contribution of each particle to the pixel:
 \begin{equation}
 \alpha_i \times \rho_i \times w(|\vec{r}_{i,j}|),
 \end{equation} where $\alpha_i$ is a geometry factor encoding the overlap of the particle with the pixel, $\rho_i$ is the nearest-neighbor density and $w$ is the kernel function. We use the commonly employed cubic spline kernel for all SMAC calculations, where
 \begin{equation}
   w(\tilde{r})\propto
    \begin{cases}
      0, & \text{if}\ r>1 \\
      2(1-\tilde{r})^{3}, & \text{if}\ 1/2 \leq \tilde{r} \leq 1 \\
      1-6\tilde{r}^2+6\tilde{r}^3  & \text{if}\ 0 \leq \tilde{r} \leq 1/2
    \end{cases},
 \end{equation}
 where $\tilde{r}=\frac{r_{i,j}}{r_{\mathrm{hsml}}}$ and $r_{\mathrm{hsml}}$, the neighbour-density based smoothing length is a simulation output.
 The net flow through the \textit{i}-th pixel is then the sum over all particle velocity contributions to that pixel, weighted with the respective particle density contribution. The total net flow is simply the sum over all pixels. For each net flow map we also generated pure inflow and pure outflow maps in order to identify the regions of in- and outflow and analyse the components separately. Additionally, temperature and metallicity weight maps are generated in order to select for hot stream components and processed matter streams. Here the metallicity of a particle is calculated as the mass fraction sum of all elements heavier (Z) than Helium (Y) divided by the particles hydrogen mass (X) $m_Z/m_H=(1-\mathrm{X}-\mathrm{Y})/\mathrm{X}$. We then convert this value to solar metallicity, assuming the mass fractions from \citet{wiersma2009}.
 \subsection{Discerning in- and outflows}
 To identify regions of inflow and outflow, we use two approaches that are outlined in the following. 
 \subsubsection{Pixel-based philosophy}
 Whenever referred to as pixel-based flows, pixels are selected by the projection of the mass-weighted net velocity vector onto the radial direction
 \begin{equation}
 \Phi_i=\vec{v_i} \cdot \hat{\vec{r_i}}
 \end{equation}
 through each pixel, where $\vec{v_i}$ is obtained with the kernel (and geometry)-weighted sum over all particle contributions in the pixel and $\hat{\vec{r_i}}$ is the local normal radius vector. This is useful for determining the spatial configuration of the vector field and selecting inflow and outflow dominated pixels, respectively.
 \subsubsection{Particle-based philosophy}
 The global quantities are computed by summing up particles with positive and negative radial velocity separately, generating pure inflow and pure outflow maps.  \begin{figure*}[!htbp]
\centering
\subfloat[$1.2 r_{\mathrm{vir}}$]{
  \includegraphics[width=\textwidth]{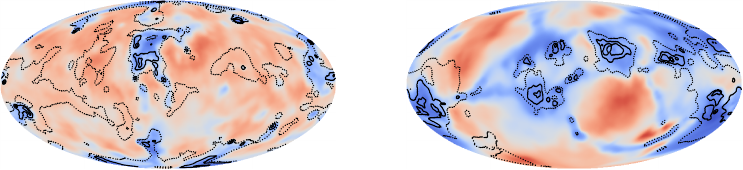}
}
\hspace{0mm}
\subfloat[$2.0 r_{\mathrm{vir}}$]{
  \includegraphics[width=\textwidth]{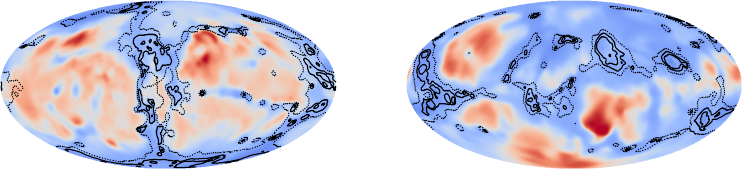}
}
\hspace{0mm}
\subfloat[$3.0 r_{\mathrm{vir}}$]{   
  \includegraphics[width=\textwidth]{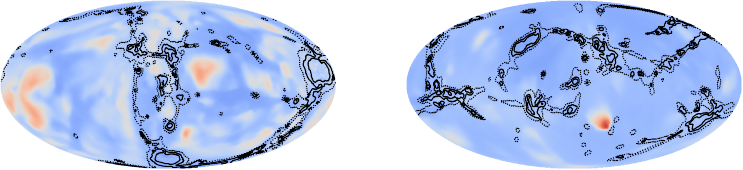}
  
}
\hspace{0mm}
\includegraphics[width=\columnwidth]{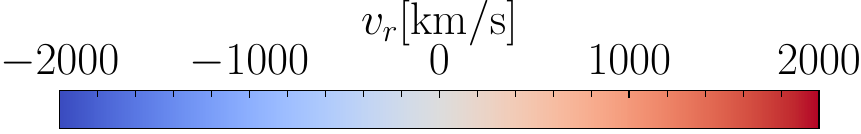}
\caption{Exemplary total mass flow map generated with SMAC \citep{dolag2005a} for Cluster 05 (left column) and Cluster 20 (right column) in the sample. Red marks the outflow dominated pixels while the inflow dominated pixels are blue. Contours again mark the overdensity at the 1-$\delta$ and the 8-$\delta$ level.}
\label{molly2}
\end{figure*}

 \subsection{Theoretical accretion model}
The fundamental theory of hierarchical structure formation can provide a theoretical first order estimate of the accretion flows one would expect in the vicinity of a halo with given mass.
 This work uses the relation for the universal mass accretion history (MAH) found by \citet{vandenbosch2002} to compute a theoretical relationship between the halo mass and the expected accretion rate onto a halo of the given mass. Based on the excursion set extension of standard Press-Schechter theory, they construct randomly sampled mass accretion histories, which then are fit by the universal relation:
 \begin{equation}
     \log\langle \Psi(M_{0},z) \rangle=-0.301\bigg[\frac{\log(1+z)}{\log(1+z_f)}\bigg]^\nu,
 \end{equation}
 where $M_0$ is the mass of the halo today (at $z=0$) and $\Psi$ is the MAH (ratio of $M$ to $M_0$) as a function of $M_0$ and redshift.
 The mass accretion rate can then be obtained by computing the derivative of this MAH with regard to time:
 \begin{equation}
     \dot{M}=M_0\times \frac{d}{dt}\langle \Psi(M_{0},z) \rangle.
 \end{equation}

 The parameters $z_f$ and $\nu$ are free fitting parameters in this model, but \citet{vandenbosch2002} offers a recipe to directly compute them from cosmology and mass scale. Since from a simulation perspective all information about the cosmology is present, this prescription is used to calculate the curve.
 It should be noted that the \citet{vandenbosch2002} approach is a backwards one: It calculates the MAH starting from the properties of the haloes today and reconstructs the assembly working towards higher redshifts. To compare the inflow of a general mass bin to the theoretical relation we therefore extrapolated the bin mass at $z=0.25$ to an expected present-day mass using the linear growth factor 
 \begin{equation}
   M_{\mathrm{bin},0}=\frac{M_{\mathrm{bin},z=0.25}D_0}{D(z=0.25)}.
   \label{nofit}
 \end{equation} 
 In this work $D_0=D(z=0)$ is calculated using the approximate formula found by \citet{carroll1992} for simplicity
 \begin{equation}
    g(\Omega_\Lambda,\Omega_M)=D/a\approx \frac{5}{2}\Omega_M[\Omega_M^{4/7}-\Omega_\Lambda+(1+\frac{\Omega_M}{2})(1+\frac{\Omega_\Lambda}{70})]^{-1}
 \end{equation}
 We also employ an alternative approach, where the relationship between the halo mass at $z=0.25$ and the halo mass today is left as a free-fitting parameter, that is parametrising
 \begin{equation}
     \dot{M}=\alpha \times M_\mathrm{bin}\times \frac{d}{dt}\langle \Psi(\alpha \times M_{bin},z) \rangle.
     \label{fit}
 \end{equation} The two methods are compared in the following section.
 
 \section{Characterising the mass flows in the outskirts of Magneticum haloes} \label{Sec:Flows}

 In most conventional structure formation models, haloes are assumed to grow self-similarly and bottom-up with the additional assumption of spherical collapse as the main mechanism in the non-linear growth regime. Examples for this are the earlier described semi analytical MAH model from \citet{vandenbosch2002} as well as the model from \citet{dekel2009}, both of which are built on the extended Press-Schechter formalism (EPS). However, the now well-studied cosmic web is already a manifestation of asphericities and non-linear effects arising in the large scale regime \citep{hidding2014}, posing the questions whether accretion models based on sphericity and self-similarity hold in realistic cosmic environments such as fully hydrodynamical cosmological simulations, and whether self-similarity itself might already be broken at the large-scale flow level. 
 \Cref{molly} shows the density, velocity, temperature and metallicity maps resulting from the scheme described in Section 2.5 for an example halo with $M_{\mathrm{vir}}=1.61 \times 10^{15}M_\odot$ (cluster 05). To illustrate the correlation between density and flow, temperature and metallicity, density contours in each panel indicate the regions with mean density and a factor of 3 and 8 times the mean density in the respective maps. As can clearly be seen, we find dense regions to be dominated by inflow and lower temperature gas indicating a cold filamentary inflow. In contrast underdense regions of higher temperature indicate an isotropic outflow most effectively expanding into underdense regions. To investigate this more closely, we split the maps into their inflowing and outflowing components and analyse them separately over a large range of radii and halo masses. For this we compute the instantaneous in- and outflow at five spherical radii ($r\in [1.2,2.0,3.0,4.0,5.0] \times r_{\mathrm{vir}} $) of increasing radius. An example of such a computation can be seen in \cref{molly2} for two galaxy clusters from the $M_{\mathrm{vir}}>10^{15}M_\odot$ bin, the 5-th (left columnm, also shown in \cref{molly}) and 20-th (right column) most massive cluster in {\it Box2b/hr} respectively (hereafter referred to as cluster 5 and cluster 20) for the full radial range. As can be seen, the inflow (blue colour) follows the higher density regions radially, while the outflows (red colour) are distributed more isotropically across the underdense regions. For both haloes, the outflowing component fades at larger radii.
\begin{figure*}[!htbp]
     \includegraphics[width=\textwidth]{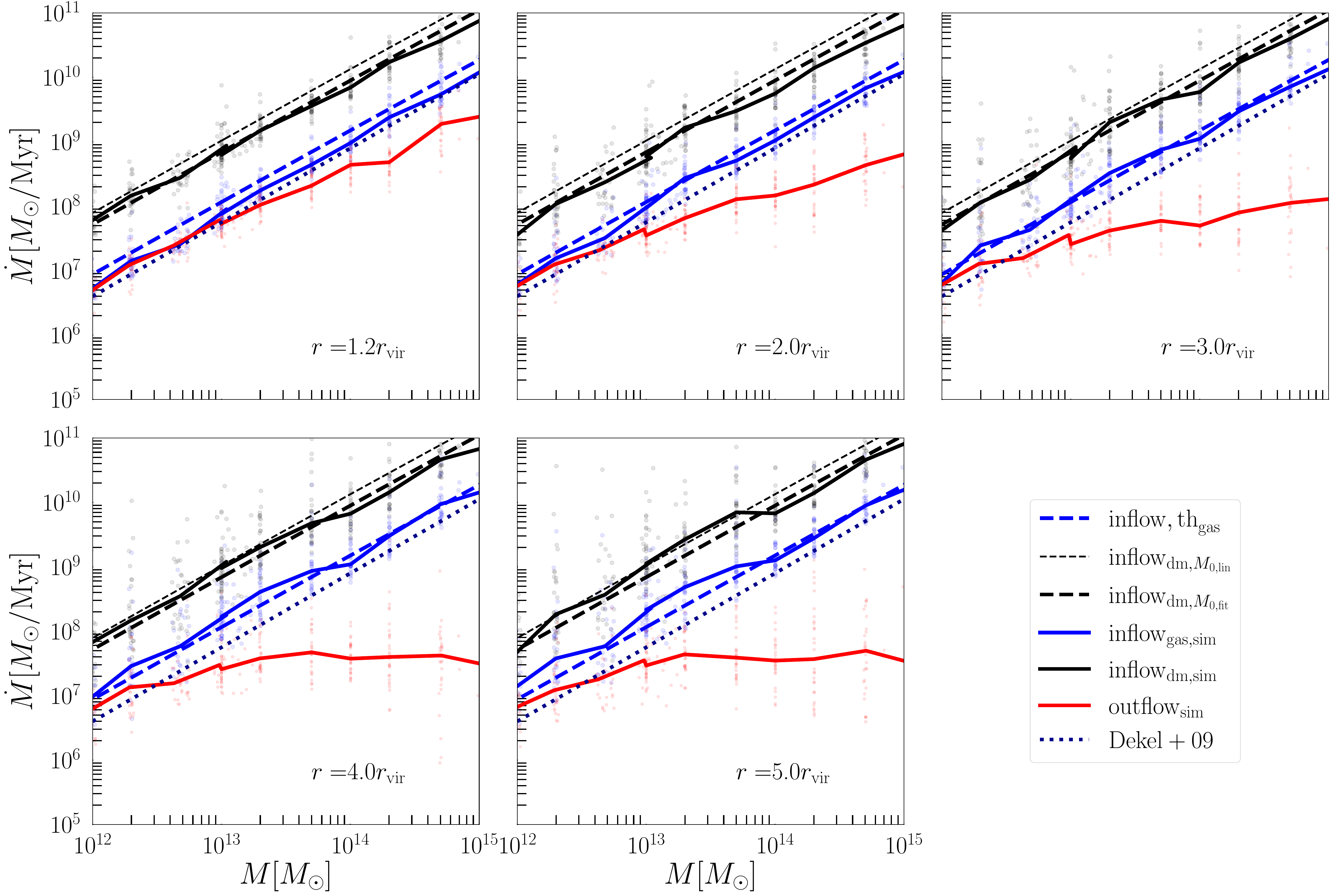}
    \caption{Comparison of the total in (black median line ($\mathrm{inflow}_\mathrm{dm,sim}$) for dm, blue median line for gas ($\mathrm{inflow}_\mathrm{gas,sim}$), solid lines)- and outflow (red solid median line ($\mathrm{outflow}_\mathrm{gas,sim}$), gas only) in the Magneticum simulation to the \citet{vandenbosch2002} model (dashed black lines, thin for the linearly extrapolated $M_0$ ($\mathrm{inflow}_\mathrm{dm,M_{0,lin}}$) , \cref{nofit}, bold for the fit $M_0$ ($\mathrm{inflow}_\mathrm{dm,M_{0,fit}}$), \cref{fit}). The inflow component is decomposed into a DM accretion flow and a baryonic accretion flow using the baryon fraction and the \citet{andreon2010} model to compensate for the increasing stellar component at low masses. Additionally we show the gas accretion model presented by \citet{dekel2009} as the dark blue dotted line. All flows are shown at 1.2,2,3,4 and 5 virial radii (different panels).}
    \label{bins}
\end{figure*}

\subsection{Accretion}
To obtain a more statistical view of these observations, we compute the surface based in- and outflows for the complete sample (from $M_{\mathrm{vir}}=10^{12}-10^{15}M_{\odot}$). \Cref{bins} shows the flow rates versus halo mass at the five radial bins. The blue line in \cref{bins} shows the surface-based total gas inflow of the sample while the black solid line shows the DM accretion. For the gas inflow, we find an overall trend of rates increasing with halo mass that is consistent across the complete radial range from 1.2 to 5.0 $r_{\mathrm{vir}}$. This clearly demonstrates gas accretion to be a self-similar process from Milky Way-mass haloes to the most massive galaxy clusters, implicating halo internal processes to be the driver of the break in self-similarity, not the accretion flows. The comparison of dark matter (DM) inflow rates with the theoretical curves obtained from \cref{nofit} (dashed black, bold) demonstrates an over-estimation of accretion from dark-matter-only linear theory relative to full hydrodynamics. The overall shape of the linear infall scaling relation is nonetheless recovered well from the simulations. Strikingly, leaving the uncertainty about final halo masses to a fitting parameter according to \cref{fit} (dashed black) achieves an even better agreement between the DM infall relations, as the normalisation brings the curves to closer agreement with each other. This is remarkable when considering the simplicity of the underlying assumptions: The calculation of \citet{vandenbosch2002} is based on a purely linear evolution of the DM density perturbation field, while the haloes considered here have significant amounts of gas and lie firmly in the non-linear regime. Additionally, as \cref{molly2} shows, the infall is geometrically far from the spherical symmetry of the top-hat collapse model that the upcrossing thresholds of the excursion set formalism are obtained from \citep[e.g.][]{navarro1997}, but rather is organised into filamentary structures that dominate the overall accretion flow. Comparing the inflowing gas curves (blue) of \cref{bins} to the DM curves shows that even close to the virial surface of the halo the infalling gas is still tightly coupled to the DM accompanying it. Only upon encountering the pressure supported, hotter gas closer to the halo centre at the shock surfaces (for a further discussion on where these surfaces are located, see \cref{Sec:Shocks}) that especially massive haloes drive in this region will this gas uncouple from the large-scale flow \citep{shi2016}. The impact of this heating on the accretion flow can be seen in comparing the theoretical gas accretion rate (dashed blue line), obtained by multiplying the DM inflow with the cosmic baryon fraction, to the measured gas flow across the five radial bins. Purely gravitationally (as derived from the accretion model) the expected gas accretion rate is significantly higher then the actual accretion achieved taking hydrodynamics into account. From $1.2-3r_{\mathrm{vir}}$, this discrepancy is consistent with mass. Beyond $3r_{\mathrm{vir}}$
 the less massive haloes ($M<2\times10^{13}M_{\odot}$) reach the theoretically predicted gas inflow rate, while the discrepancy in the high-mass bins persists to larger radii. This effect is visible in both the DM and the gas component. However, as can be seen by the thin dashed line in \cref{bins}, this discrepancy can be almost entirely attributed to the linear extrapolation, leading to an overshooting in the theoretical estimate. Leaving the scaling as a fitting parameter (as described in \cref{Sec:Methods}) leads to much better agreement.

 \subsection{Outflow}
 \begin{figure}[!htbp]
     \includegraphics[width=\columnwidth]{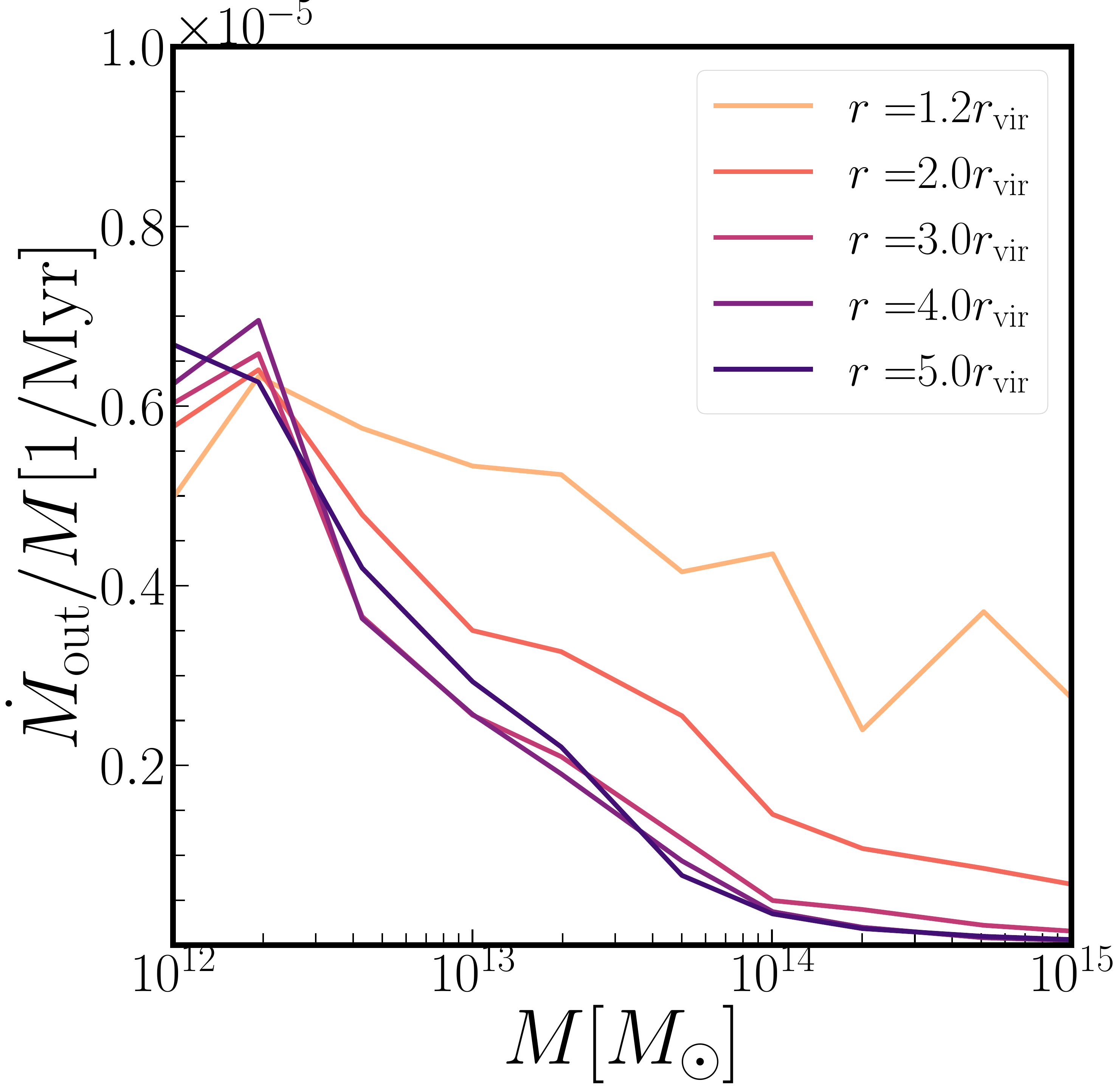}
    \caption{Relative outflow strength (the outflow rate divided by the respective halo mass, $\frac{\dot{M}}{M}$) as a function of mass. The line colours indicate the different radii the flow strength is measured at.}
    \label{binsout}
\end{figure}
 There are two main channels that can drive outflowing motion in the outskirts of an assembling halo, namely the conversion of gravitational energy from mergers and accretion to thermal energy of the gas via virialization processes and heating from AGN and star formation in the galaxies present in the haloes. The outflowing component is shown in red in \cref{bins}. In contrast to the accretion rate, the outflowing component flattens out with increasing radius, being almost completely flat at 5 $r_{\mathrm{vir}}$, breaking the self-similarity.
 \Cref{binsout} additionally shows the relative outflow strength (mass flow rate normalized to the halo mass) for each mass bin at the five radii. The breakdown of strong outflows at the high mass end can be seen even more pronounced in this figure, especially at large radii.
 
 This is most likely a manifestation of the different effectiveness of feedback processes at different scales. At galaxy scales, processes like stellar feedback and AGN feedback can kick gas out to several virial radii, maintaining a high outflow rate to larger relative distances, while in galaxy clusters and massive groups the deeper potential well can balance out the impact of heating. Note that numerical simulations employing simplified AGN feedback schemes can additionally over-enhance the impact of the (in the case of Magneticum purely thermal) feedback at galaxy to group scales, which we will discuss further in \cref{Sec:Phase}.
 
 Because of the contrast between the thermal nature of the outflow and the gravitational nature of the inflow, one would also expect a difference in the spatial distribution of the two gas components with the outflows being more isotropically distributed than the inflows due to their thermal origin. The map-making scheme described in \Cref{Sec:Methods} provides the data to verify this for the numerically simulated haloes, which we will discuss in further detail in the following section.
 \subsection{Flow geometry} \label{fgeo}
\begin{figure*}[!htbp]
\centering
\includegraphics[width=\columnwidth]{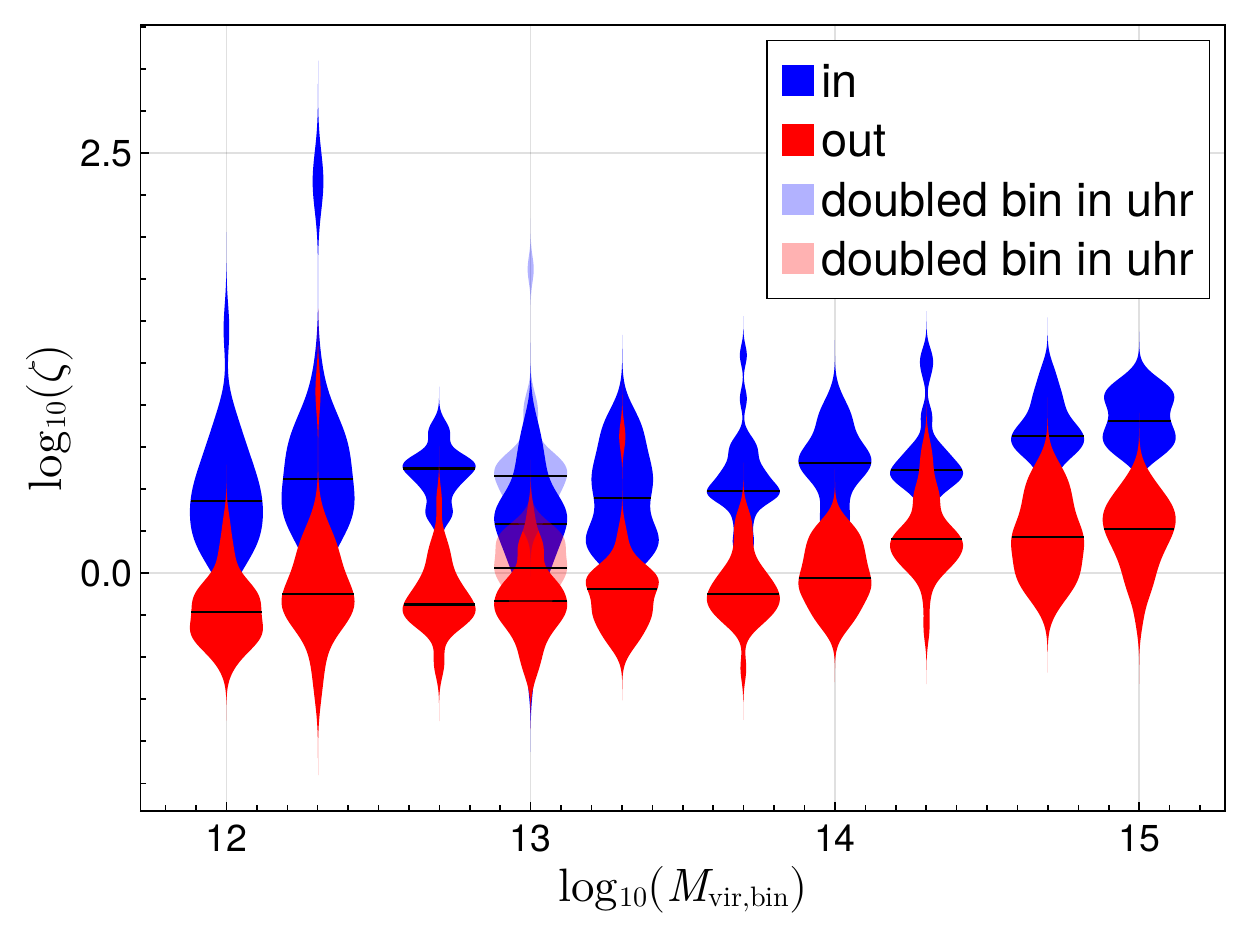}
\includegraphics[width=\columnwidth]{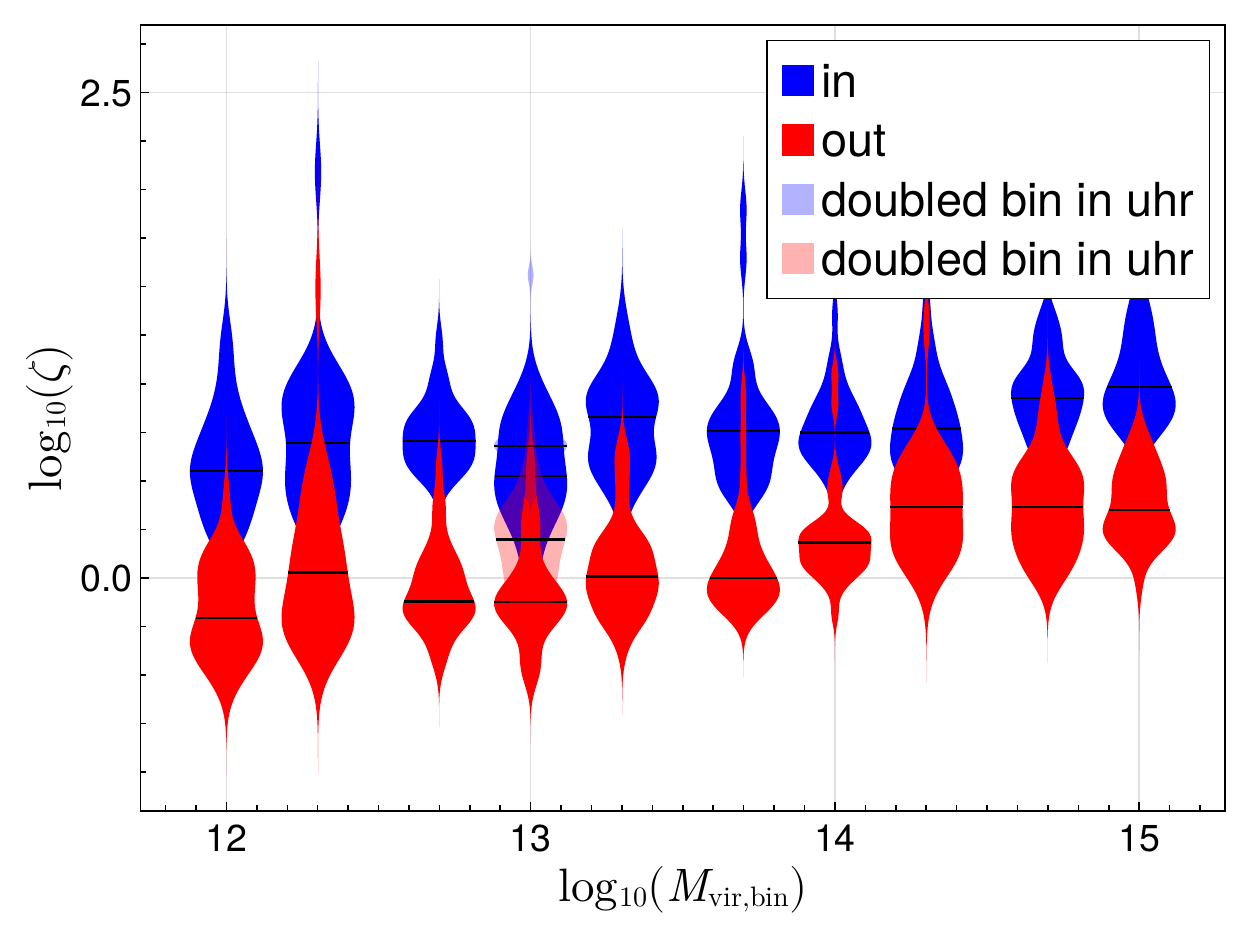}
\caption{Aspherical excess of the flows parametrized by the asphericity parameter $\zeta$ at $1.2r_{\mathrm{vir}}$ (left) and $2.0r_{\mathrm{vir}}$ (right) for the in- and out-flowing components, respectively. The violins show the vertical distribution, while the black horizontal lines show the median for each bin. The bin present in both simulations ({\it Box2b/hr} and {\it Box4/uhr}) is plotted twice to indicate the transition in resolution. }
\label{power}
\end{figure*}
 In \Cref{molly2} we showed the surface distribution of in- and outflow dominated pixels for two example haloes. It is apparent that  inflowing gas is generally more concentrated than outflowing gas, consistent with the picture of a filamentary inflow driven by gravity versus thermally driven outflows. To obtain a more statistical measure of the geometrical properties of the gas distribution (or any distribution of a scalar quantity) the flow field can be reduced to a power spectrum by computing its \textit{harmonic decomposition}. Comparable techniques have been used in the past to identify preferred directions of infall for satellite galaxies \citep{libeskind2011}, to measure the degree of filamentary to isotropic accretion in isolated cluster simulations \citep{valles-perez2020}, and to study the correlation of the 2D-azimuthal symmetries in gas and DM to various halo properties \citep{gouin2022}. We calculate the \textit{aspherical excess} on the sphere, following closely the approach by \citet{gouin2022}. This excess is given by the sum of all multipole moments higher than the monopole normalized to the monopole power, truncating the sum at the 9-th order to ideally capture the large-scale anisotropic inflow modes:
 \begin{equation}
    \zeta=\frac{1}{C_0}\sum_{l=1}^{9} (2l+1)C_l.
\end{equation}
A more spherically homogeneous field has less power in the non-isotropic angular modes, thus lowering $\zeta$. A more detailed study on this parameter, the \textbf{Asphericity $\zeta$} will be presented by Seidel et al. (in prep.)

 \Cref{power} shows the distribution of the asphericity for each mass bin at $1.2r_{\mathrm{vir}}$ (left panel) and  $2r_{\mathrm{vir}}$ (right panel). In agreement with expectations, the distribution of inflowing gas is consistently more aspherical than the outflowing component. Additionally, the inflow asphericity exhibits a wider spread between systems. This clearly demonstrates the different mechanisms at play for the in- and outflowing component. While gravity organises the inflows into filaments according to the local environment of a halo, thus exhibiting a greater variance, the thermal processes that drive the outflows are more isotropic in nature and thus show less variance between haloes in different environments. The stability of this discrepancy across the mass range indicates, that groups and galaxies are indeed also connected to filamentary flows, similar to their more massive counterparts. This reflects the fractal nature of the cosmic web \citep{gaite2019}, providing connectivity for galaxies and groups via low mass filaments. 

Only a slight positive mass trend is apparent, indicating that the flow geometry outside of the virial radius is similar across the mass scales. Especially the lack of a low-mass end enhancement is striking as this implies that the outflows generated by baryonic feedback processes do not disturb the homogeneity of the hot gas outside of the virial surface even for small groups. This result is remarkable because feedback strength in a galaxy varies on small spatial scales, so one would expect more power in the high-frequency modes in the low mass bins, where these processes contribute more to the overall outflow. A potential reason for the missing signal here is the currently simplified sub-grid model for feedback in the Magneticum simulations with a distinct lack of directional kinetic feedback.
 
A discrepancy between the two simulation volumes can be seen at the mass bin with $10^{13}M_{\odot}$: In contrast to the mass flow rates, which are averaged over the entire spherical surface the asphericity is enhanced in the smaller, higher-resolution {\it Box4/uhr}. A possible explanation is that the highest mass bin of {\it Box4/uhr} by construction contains some of the largest collapsed haloes in the box, which tend to be very active and rather atypical in small volume cosmological simulations, so the bin value might simply not be representative for this mass bin. 
An additional possible factor is the difference of resolution, leading to a better resolved web structure around the haloes in the smaller {\it Box4/uhr}. This can be understood by considering how the total mass is split into discrete particles: In the higher-resolution box, particles have a smaller characteristic scale, which, when keeping the physical scale constant, will produce a fluctuation of the flow field with a smaller minimal scale than in the low-resolution case.   

\subsection{Redshift evolution of the measured flows}
   \begin{figure*}
  \begin{center}
	\includegraphics[width=\textwidth]{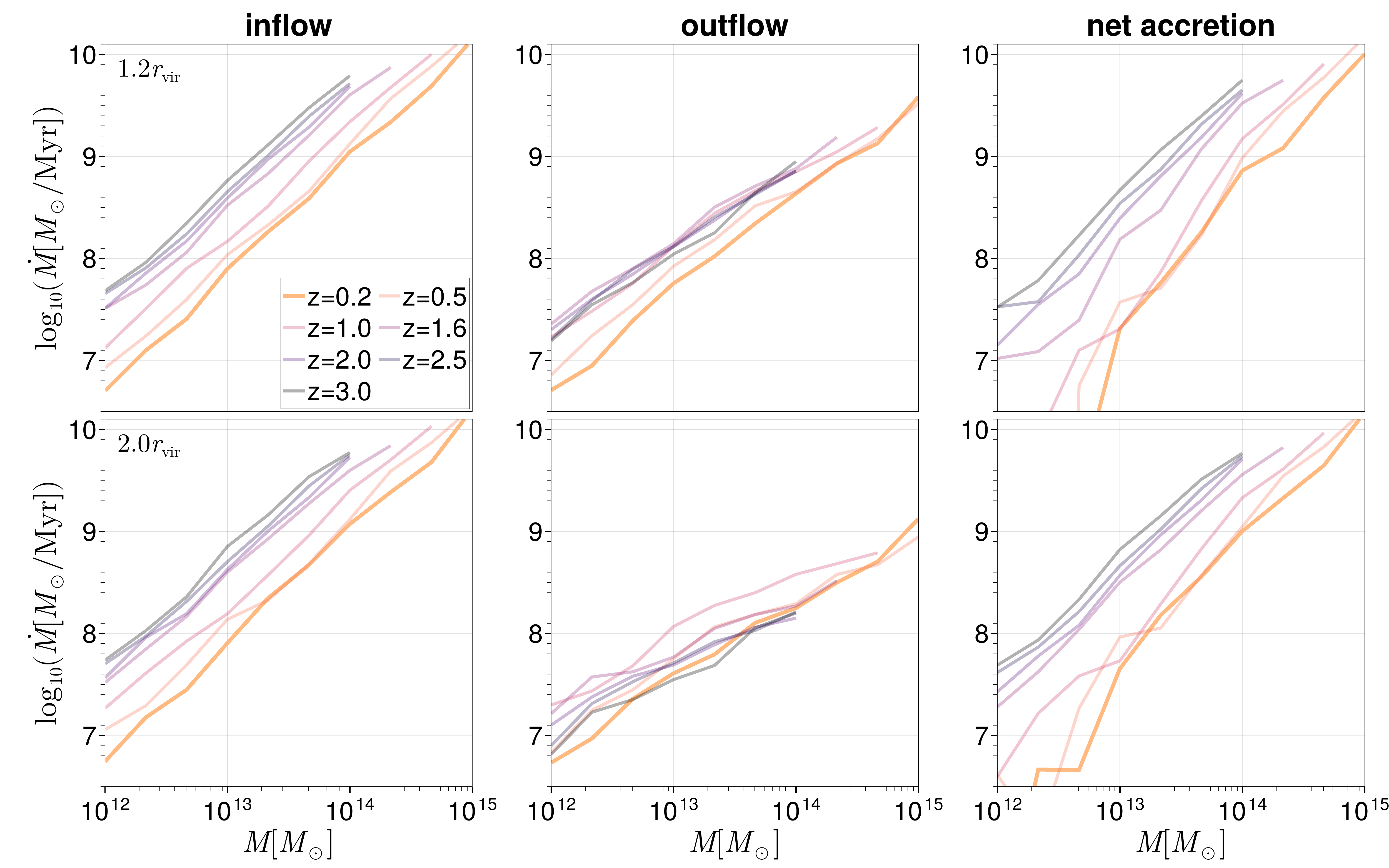}
    \caption{Redshift evolution of in-, out- and net flow (inflow $-$ outflow) at $1.2r_{vir}$ (top row) and at $2.0r_{vir}$ (bottom row). The bin masses (x-axis) are identical to the ones used in \cref{bins}. The colours indicate the redshift of the curve.}
    \label{timevo1}
    \end{center}
\end{figure*}
 \begin{figure*}
    \centering
	\includegraphics[width=\textwidth]{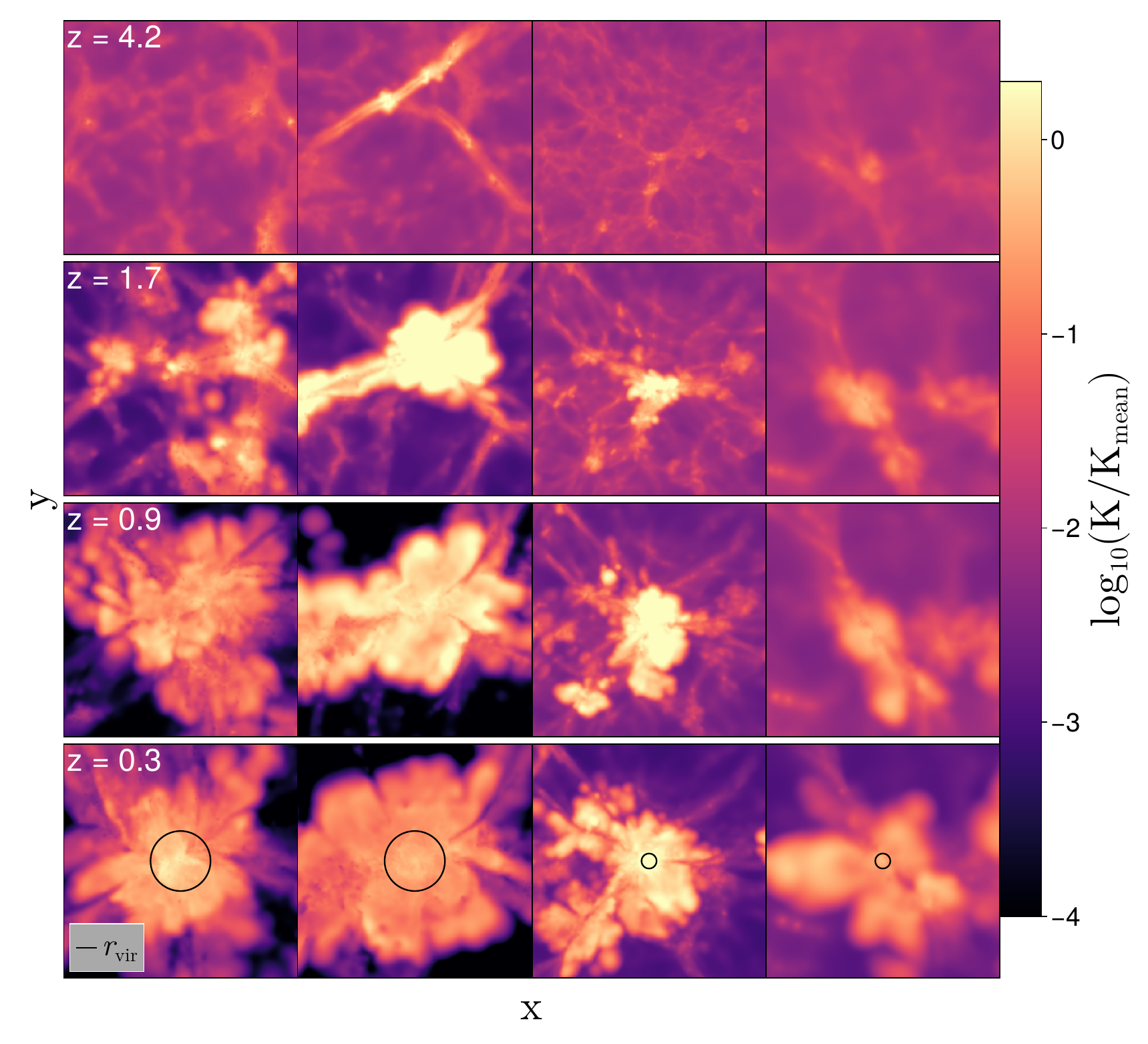}
    \caption{Entropy evolution of a $1.3\times10^{15}M_{\odot}$ mass non cool-core cluster, a $1.6\times10^{15}M_{\odot}$ cool-core cluster, a $1.6\times10^{13}M_{\odot}$ size group and a $0.91\times10^{12}M_{\odot}$ halo from left to right. The black circle indicates the virial radius of each object. The box lengths are scaled to 10 times the virial radius for the clusters and 40 times the virial radius for the groups/galaxies with a slice thickness of 1 Mpc/h for all images.}
    \label{shocks}
\end{figure*}
\begin{figure*}
\centering
\includegraphics[width=\textwidth]{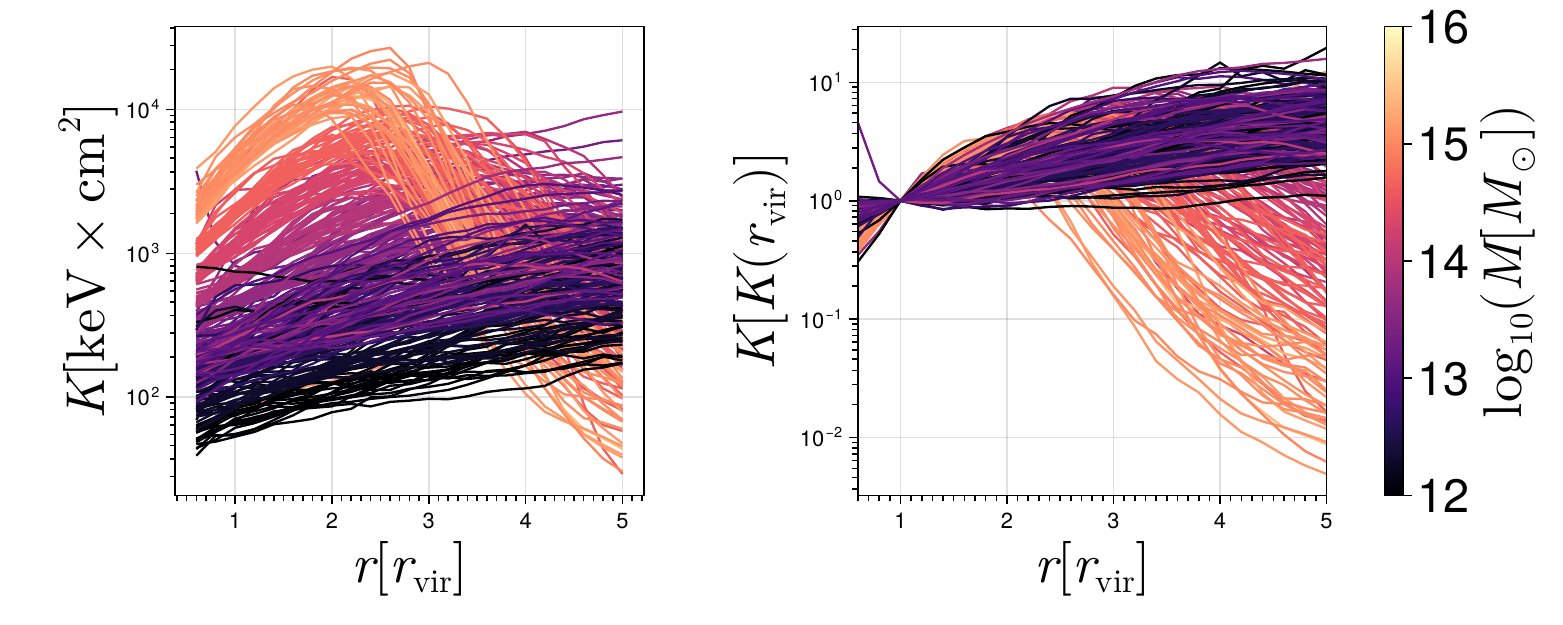}
\caption{Absolute (left panel) and scaled ( $K(r)/K(r_{\mathrm{vir}})$ right panel) radial entropy profiles for all haloes in the sample. The virial mass is indicated by the colour of the curve.}
\label{entropy}
\end{figure*}

\begin{figure}
\centering
\includegraphics[width=0.48\textwidth]{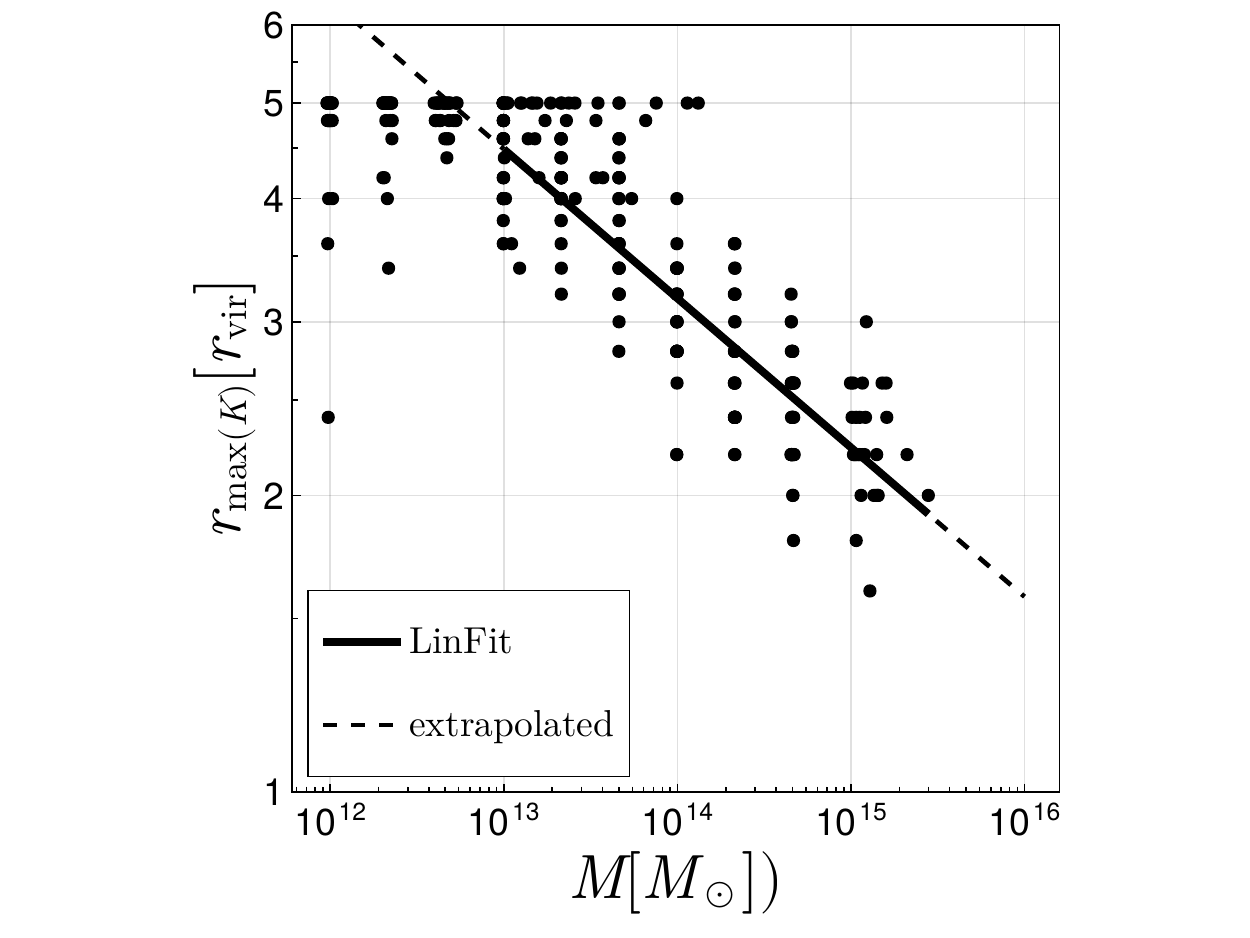}
\caption{Mass dependence of the radius of maximal entropy. The dots represent the individual haloes of the sample. The line shows a Linear fit in log-log space obtained from the haloes in the range where there is a clear entropy peak visible. The obtained parameters are $t=2.6193\pm0.1157$ and $m=-0.1513\pm0.0008$ using a $\mathrm{log}(r_{\mathrm{max}(K)})=m\times\mathrm{log}(M/M_\odot)+t$ model.}
\label{Peak}
\end{figure}
In the hierarchical picture of structure formation haloes form bottom-up with low mass haloes forming early and then collapsing into larger structures at late times \citep{press1974}. As the cosmic web assembles we thus expect the accretion flow rates of our sample of isolated haloes to decrease with time overall.
We consider the redshift evolution of the flows:
\cref{timevo1} shows the inflow (left column) outflow (middle column) and the net accretion rate as a function of mass at different times, with the colour of the curves indicating the redshifts across all panels. We find that the inflow rates at all halo masses decrease self similarly with time. In contrast, the outflow is generally more stable over time than the inflow, showing an overall decrease by a factor of 2-4 vs. a factor of 5-8 for the inflow. Both flow modes vary mostly self-similarly across the entire mass range with decreasing redshift, conserving the slope of the flow-mass relationships. The consequence of relatively stronger outflows at the low mass end, which was discussed in the previous section, can be seen in the rightmost panel: While all haloes experience a net inflow, its strength is proportional to halo mass, with the low-mass end exhibiting significantly lower net inflow than the massive clusters, which breaks the self-similarity in the net accretion rate. With decreasing redshift this relationship steepens, caused by a slight increase in the slope of the outflow-mass relationship, further highlighting the breakdown of self-similarity between the outflows in low and high mass haloes. Cluster-size objects at late times seem to be able to compensate the relatively weaker impact of thermal feedback more effectively at late times, steepening the outflow mass relationship.
 
One possible explanation for this is that at low redshifts most cluster-sized objects have undergone a massive merger through their assembly history and thus also have more energy injected into their outskirts as well as merger shocks or even MAH shock fronts where the gas is heated rapidly \citep{zhang2020}.  As this transition between feedback driven and shock propagation driven outflows (where the latter are sourced by gravitational heating and mergers) also impacts the chemical evolution of the gas, the following section will analyse the shocks more closely before the interacting thermal and chemical properties of the pre- and post-shock gas are considered.

\section{Shocks in the outskirts of Magneticum haloes}
\label{Sec:Shocks}

Shocks play a central role in the formation of structures and the fragmentation of matter therein. In the simplest picture of a collapsing gas cloud (e.g. \citet{rees1977}) with even slight deviations from a perfectly spherical collapse, shock heating thermalises the kinetic energy and critically influences further mass accretion: Depending on the ratio of the (radiative) cooling time to the infall time $\frac{t_{\mathrm{cool}}}{t_{\mathrm{dyn}}}$, collapse might either continue in free-fall if cooling is efficient enough to diminish the pressure support from the heated gas (case I, $t_{\mathrm{cool}}<t_{\mathrm{dyn}}$) or enter a quasi-static contraction stage with a hot atmosphere (case II, $t_{\mathrm{cool}}>t_{\mathrm{dyn}}$) \citep{dekel2009}. The distinction between these regimes is determined by the overall mass of the system, with the transitional regime traditionally in the range of $M_\mathrm{vir}=10^{10}-10^{12}M_{\odot}$ \citep{rees1977}.
 \begin{figure*}
\centering
  \includegraphics[width=0.3\textwidth]{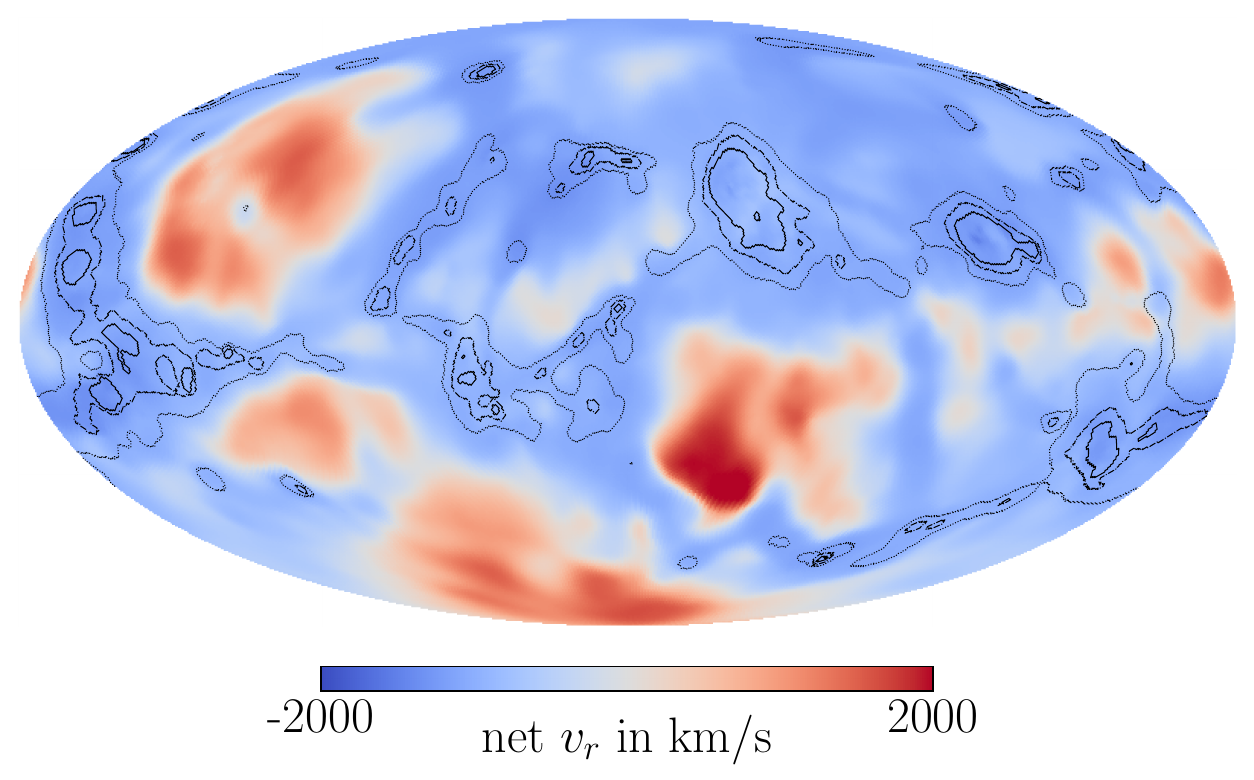}
  \includegraphics[width=0.3\textwidth]{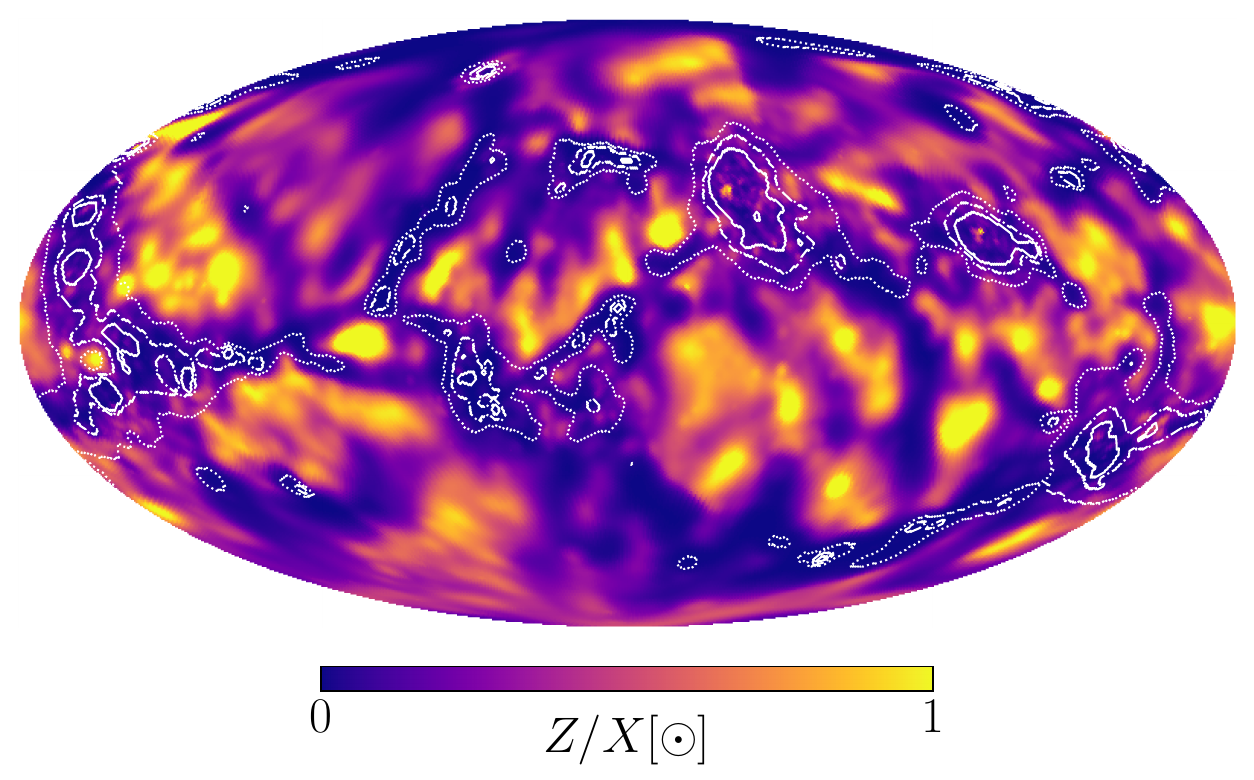}
  \includegraphics[width=0.3\textwidth]{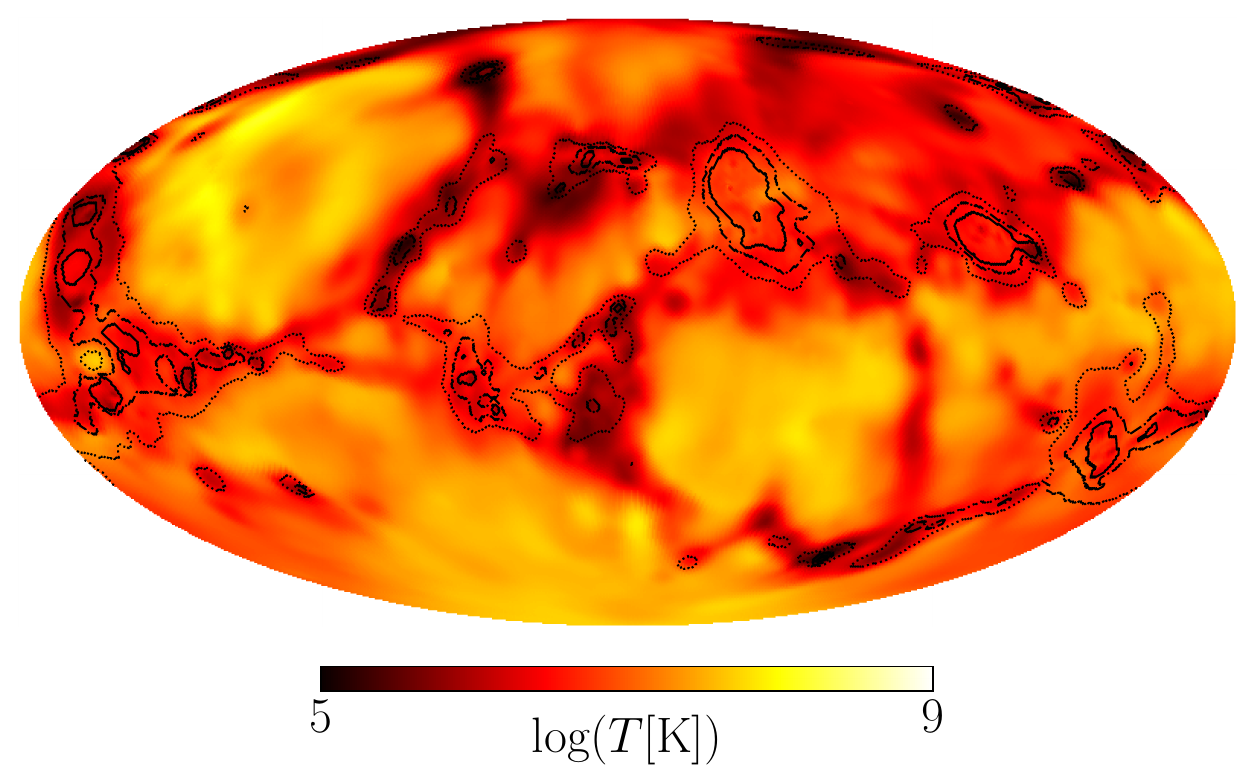}
\caption{Comparison of the gas properties (Average radial velocity (left), Metallicity (middle) and Temperature (right)) at 2.0 $r_{\mathrm{vir}}$ for cluster 20. Contours the same as in \cref{molly}}
\label{molly3}
\end{figure*}
\begin{figure*}
  \includegraphics[width=0.8\textwidth]{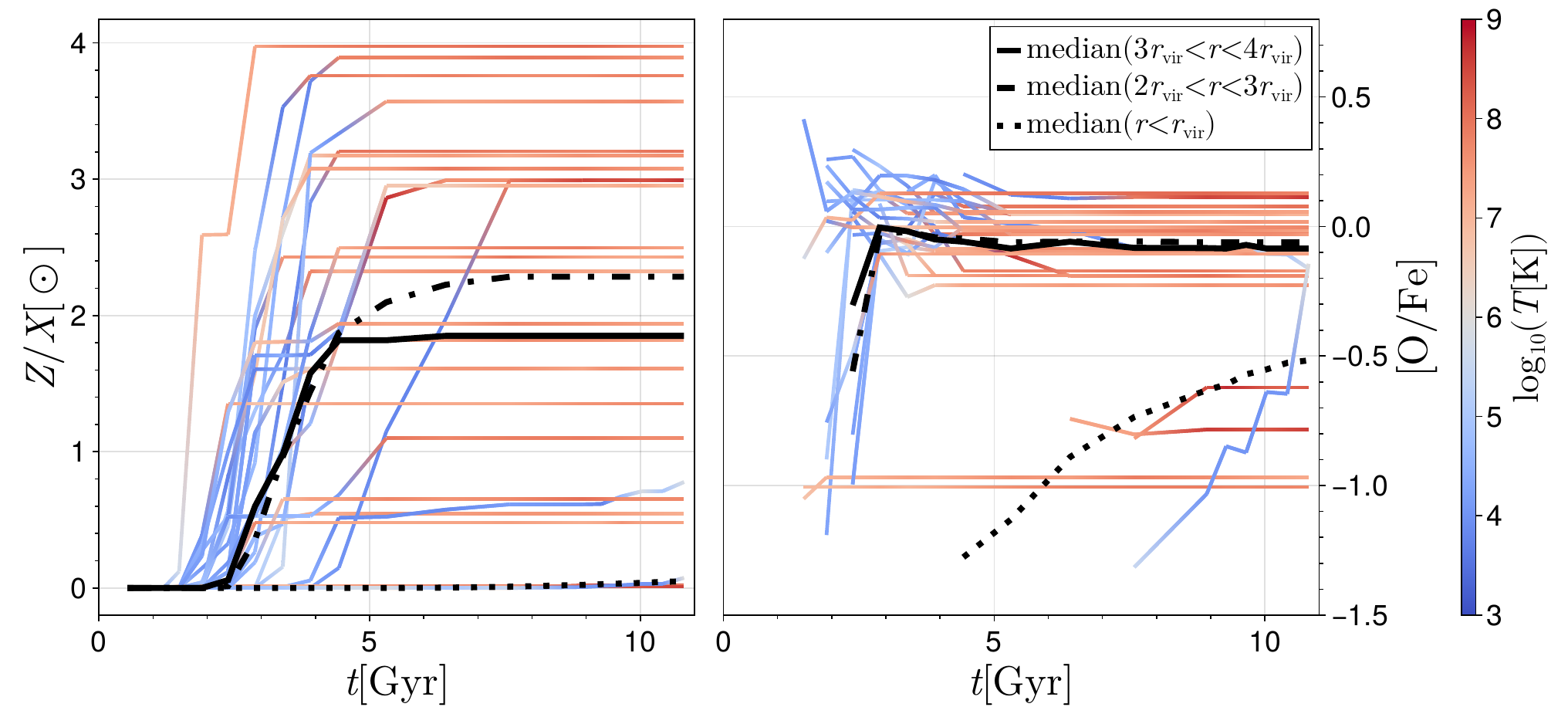}
    \hspace{0mm}
    \centering
\caption{Evolution tracks of 40 individual particles in overall metallicity (left panel) and O/Fe ratio (right panel). Particles were selected randomly between $3r_{\mathrm{vir}}$ and $4r_{\mathrm{vir}}$ with a radial peculiar velocity exceeding 100 km/s (outflowing) and a final metal mass exceeding $m_Z/m_H=10^{-4}$. The black lines show the medians of 1000 randomly selected particles each for three different radial ranges. The line colour indicates the temperature of the particle at the given scale factor.}
\label{ztrack}
\end{figure*}
\begin{figure}
\includegraphics[width=\columnwidth]{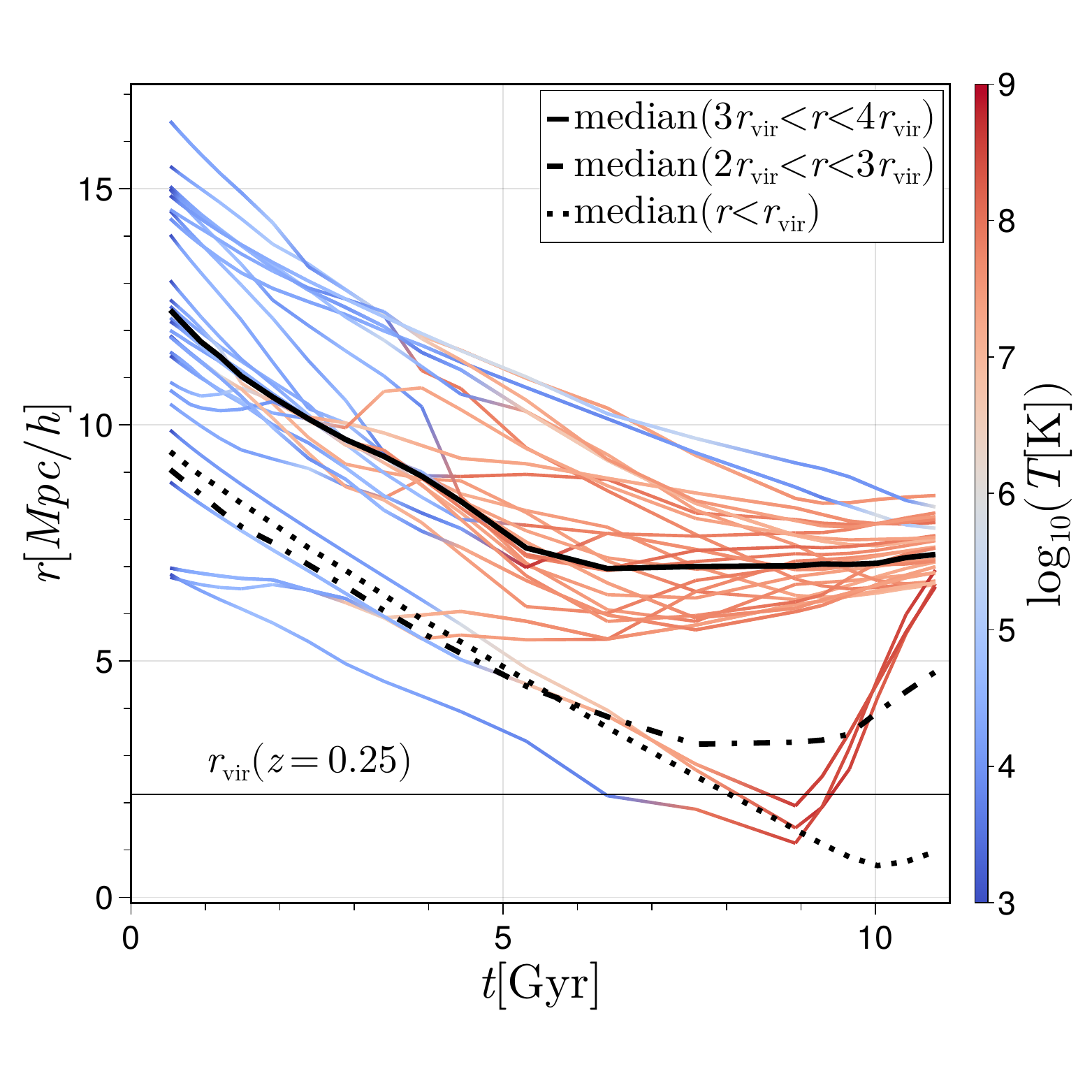}
\caption{Evolution tracks of the same 40 particles as in \cref{ztrack} in radial distance to the cluster centre. The black lines again show the medians of 1000 randomly selected particles each for three different radial ranges. The line colour indicates the temperature of the particle at the given scale factor.}
\label{rtrack}
\end{figure}
An important distinction between the two modes of accretion is the radius at which the shock stabilises. In the low-mass case (I) the stabilisation radius is deep inside the DM halo ($r_{\mathrm{shock}} \ll r_{\mathrm{vir}}$) while more massive haloes are expected to have their accretion shocks stabilise on distances of the order of the virial radius or larger $r_{\mathrm{shock}}\geq r_{\mathrm{vir}}$. 

In the complex environment of non-linear structure formation the transition between the rapid-cooling and slow-cooling accretion channels, namely the shock position, is significantly impacted by additional factors: Denser gas is organised into filaments at inflow and thus cools more efficiently and penetrates further into the halo unaffected by the shock \citep{dekel2009}. Furthermore, as previously mentioned, merger shocks interact with the accretion shocks \citep{zhang2020} in the outskirts of massive clusters. Additionally the energy injected by feedback plays a more important role at the low mass end. This affects both the inflowing and the outflowing gas: The outflowing component is heated and can thus reach larger radii in the shallower potential of low-mass haloes. On the other hand feedback can also lead to a gentle heating of the infalling gas before being shocked \citep{murante2012}, especially in low mass haloes. Hydrodynamical cosmological simulations like Magneticum therefore offer a unique laboratory to study the cluster shock dynamics in realistic environments and probe for the stability of simple mass-scaling arguments like the one outlined above.

One cannot draw precise conclusions about the position of the shocks themselves from the flow fields alone. To assess the scale dependence of shock strength and position, we calculate the pseudo-entropy ($K$), given by e.g. \citet{zhang2020} as
\begin{equation}
K=\frac{T}{{\rho}^{\gamma-1}},
\end{equation}
averaged over radial shells across the complete sample, where usually an adiabatic index $\gamma=5/3$ is assumed. \Cref{shocks} shows the entropy environment in a quadratic slice of $10r_{\mathrm{vir}}$ side length for halo 20 ($10^{15}M_\odot$) and halo 5 ($10^{15}M_\odot$) from {\it Box2b/hr} as well as halo 31 ($10^{13}M_\odot$) and halo 538 ($10^{12}M_\odot$) from {\it Box4/uhr}. Generally it can be seen from the bright entropy edges, where the entropy peaks before dropping off, that the accretion shocks stabilise at much larger relative radii for the group size and galactic haloes, aided by the shallower potential of these haloes and the thermal feedback. This can also be seen in \cref{binsout}: The outflow strength relative to the halo mass is almost monotonically decreasing as a function of mass at larger radii. In general, \cref{shocks} demonstrates that the radial peaks of entropy serve as a simplified tracer of where the outer shock surfaces lie.

We extend the radial profiles inwards to $0.6r_{\mathrm{vir}}$ in order to detect the entropy peaks if they are found at smaller halocentric distances. \Cref{entropy} shows these radial entropy profiles for all haloes in the sample in absolute entropy (left column) and scaled to the entropy at the virial radius (right column). There is a clear evolution in the shape of these profiles with mass: The shocks (i.e. the entropy peaks) are smoothed out with decreasing mass of the central halo. For the haloes that exhibit an entropy peak within the probed radial range (up to $5r_\mathrm{vir}$), it is generally broader and peaked at greater radii with decreasing mass. This is in line with the general finding of the so-called closure radius (i.e. the radius at which the baryon fraction reaches the cosmic mean value) to scale with halo mass, as observed across various simulations in the group/cluster regime \citep{angelinelli2023}. Most of the haloes that show a distinguishable peak have it in the region between $2-4 r_{\mathrm{vir}}$. This is consistent with slow cooling accretion shocks and possibly additionally merger driven outwards shock propagation. To demonstrate this more quantitatively, \cref{Peak} shows the radius of maximum entropy as a function of mass for our halo sample. We find a consistent trend of decreasing scaled radius of maximum entropy as a proxy for the approximate shock position with mass. A possible explanation is the relatively stronger potential the gas moves into these heavier haloes, allowing the hot post-shock gas to expand to relatively smaller radii overall. The general dynamics can be expected to be consistent across the total mass range. We will therefore now investigate the interaction of gas accretion and shocks specifically for a galaxy cluster from the highest mass bin.

\section{Case study: Cluster 20}
\begin{figure*}
 \begin{center}
	\includegraphics[width=0.65\textwidth]{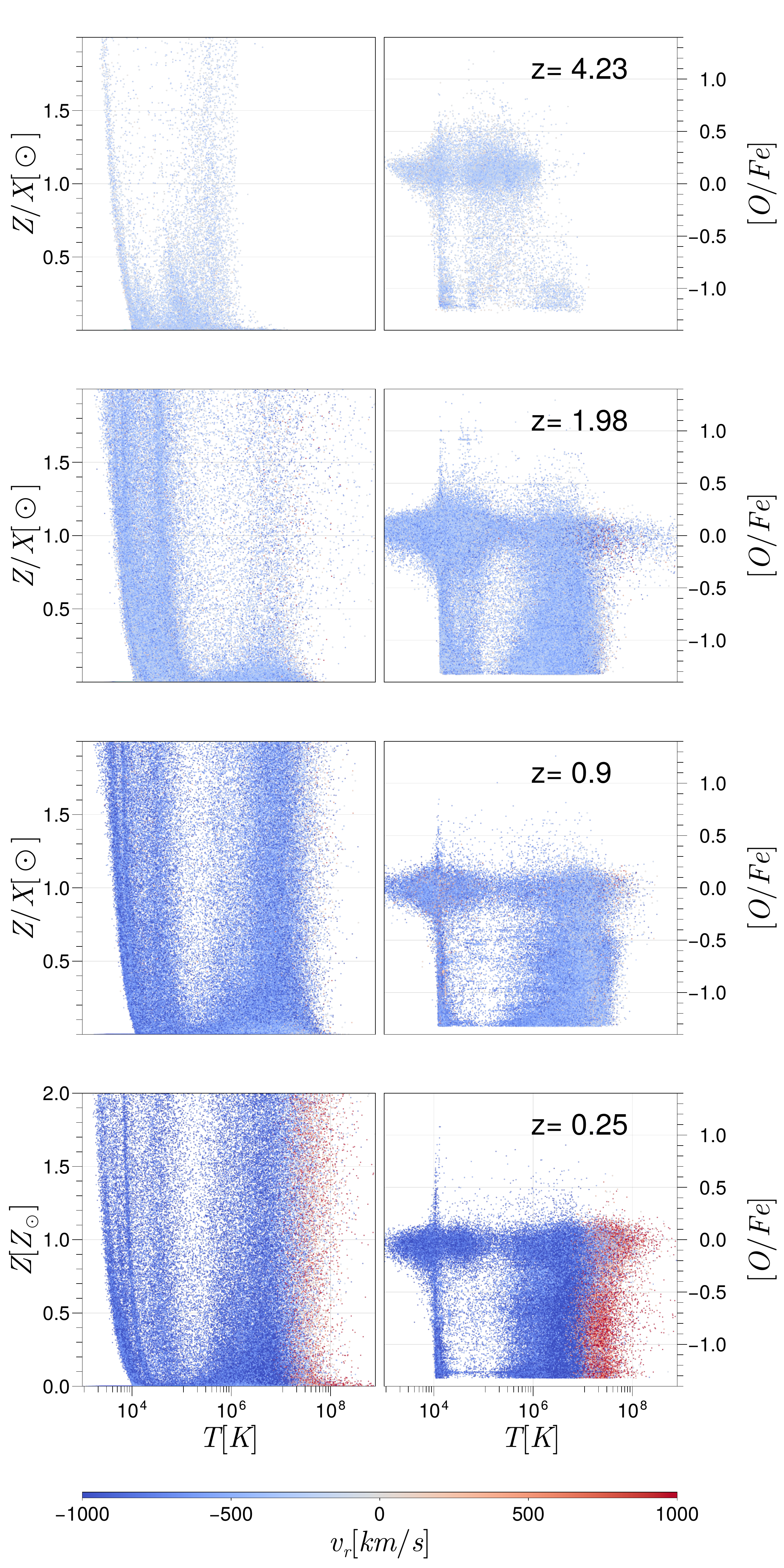}
    \caption{Particle temperature versus metallicity (left column) and oxygen/iron ratio (right column) evolution with redshift. Inflowing particles are shown in blue while outflowing particles are shown in red.}
    \label{T-Z}
 \end{center}
\end{figure*}
The thermal properties of the infalling and outflowing gas and how the two components (in- and outflow) interact at the shock boundaries is of critical importance for many processes internal to the halo, for example replenishment of gas reservoirs for star formation in galaxies or the dynamical states of galaxy clusters. Because the jumps in the thermal properties of the gas are best resolved in the largest haloes (i.e. the shocks are the most pronounced), we will now focus on the gas properties in one of the biggest clusters in the Magneticum simulations, cluster 20 which was shown to be an excellent match for the observed galaxy cluster Abell 2744 by \citet{kimmig2023}. Specifically we will focus on the infall and enrichment history of the gas that constitutes the outflowing component today. The abundance of heavy elements is routinely used in the literature to determine the origin and enrichment time of gas as well as potential mechanisms to strip it from its host galaxies in cluster environments \citep[e.g.][]{biffi2018}. The velocity, temperature and metallicity maps for this cluster are shown in \cref{molly3}. Consistent with expectations for an Abell 2744 counterpart, this cluster is fed by multiple filaments, mirroring the multiple mergers it is currently undergoing \citep[see]{kimmig2023}. We will therefore elucidate specifically the impact of the thermal structure and accretion on the metallicity of the outflowing gas. We will then assume a more global perspective on these gas properties in the final section.

The total metallicity of the gas found in the outskirts of clusters and its radial and azimuthal distribution can shed light on the overall dynamics of the infalling gas, that is where it is cold and dense enough to form stars that provide an influx of metals. This information can be further compounded by considering the enrichment type by analysing the mass ratios of different elements predominantly generated in different enrichment processes. One indicator for the type of enrichment and critically when it takes place is the ratio of oxygen to iron mass in the gas, as oxygen is the most common alpha element \citep{kobayashi:2020}. Early-type enrichment (i.e., enrichment from massive and short-lived stars on short timescales) is predominantly generated by core-collapse supernovae type II \citep[e.g.][]{nomoto2013,loewenstein2006}, which produce lighter (alpha) elements in greater abundance and thus increase the ratio of oxygen to iron mass. Significant late-type enrichment, which is generated by thermonuclear supernovae type Ia, lowers this ratio if the gas is in contact with the stellar component on longer timescales \citep{deplaa2007}. Extending the particle tracing method introduced by \citet{biffi2018} to larger radii we follow the thermal and chemical evolution of the outskirt gas as it is accreted and heated by the central cluster.

 \Cref{ztrack} shows the evolution tracks of gas particles that end up in the outflowing component of the cluster at the last snapshot ($z=0.25$) with regards to total metallicity (left panel) and O/Fe ratio (right panel). Here O/Fe is defined as $[\mathrm{O/Fe}]=\log(\frac{m_\mathrm{O}}{m_\mathrm{Fe}})-{log(\frac{m_\mathrm{O}}{m_\mathrm{Fe}})}_\odot$, using again the mass fraction from \citep{wiersma2009}. Particles were selected from three radial bins (1000 particles per bin) within $1r_{\mathrm{vir}}$, between $2r_{\mathrm{vir}}$ and $3r_{\mathrm{vir}}$ and between $3r_{\mathrm{vir}}$ and $4r_{\mathrm{vir}}$. The coloured tracks show the evolution of 40 randomly selected particles from the outermost bin with blue indicating a cold particle temperature and red a hot temperature. Both panels demonstrate enrichment in the outflowing component to happen primarily early on, while the particles are infalling or in the halo centre before they are heated. Essentially all vertical movement in the left panel (increasing overall metallicity) occurs before $t\approx 6 \mathrm{Gyr}$ and critically also before the particles are heated. Once the infalling particles are heated to the virial temperature, no new stars are formed in their surroundings as they become part of the diffuse cluster atmosphere, stopping also the influx of metals. This is also visible in the O/Fe ratio: Particles are first enriched by core-collapse supernovae, causing their oxygen mass to increase relative to their iron mass. At later times, type Ia supernovae decrease this ratio, but this process is eventually stopped once the particles are heated completely, leaving the particles stable at relatively high oxygen to iron ratios. A second population of particles, with a lower final oxygen to iron ratio can be seen in both panels. These particles are heated early on, remain at low metallicities and undergo almost no type Ia enrichment. Comparing the median lines at different radii, the difference between the enrichment processes in- and outside the virial surface become apparent: While the two radial regions in the outskirts are enriched early and settle on a high oxygen-to-iron ratio \citep[see also][]{biffi2018}, the gas particles inside the virial radius get enriched much later and do not reach the same O/Fe ratios due to metal influx predominantly from thermonuclear supernovae. 

  \begin{figure*}
	\includegraphics[width=\textwidth]{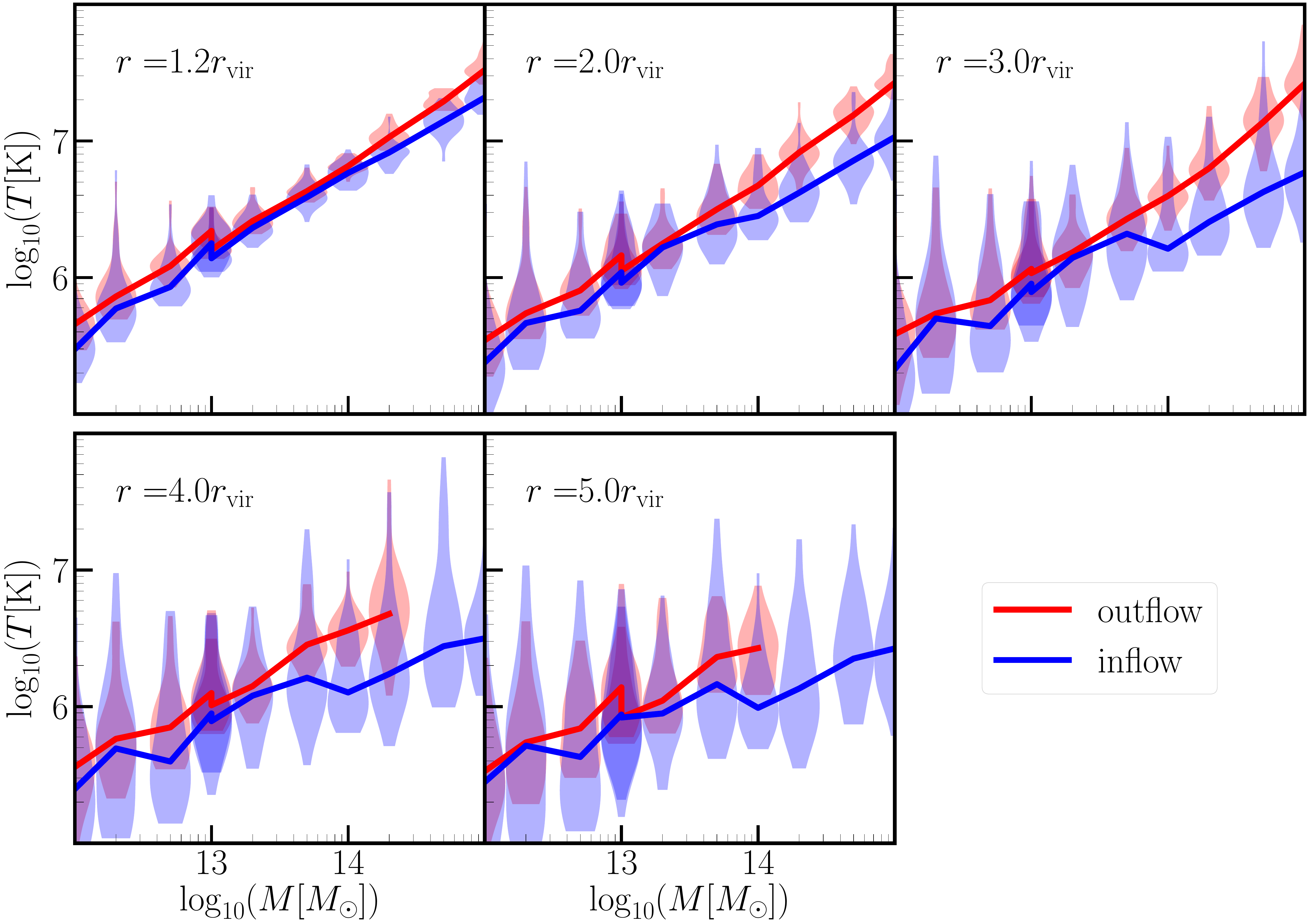}
    \caption{Temperature of net in- and outflow as a function of mass at five radii for the full halo sample. The spread of each bin is indicated by the red and blue vertical distributions. The cutoff at $10^{14} M_{\odot}$ for $r\geq 4.0 r_{\mathrm{vir}}$ stems from the fact that there are no pixels with $\textit{net}$ outflow at this scale and does not indicate zero outflow or zero temperature.}
    \label{bins3}
\end{figure*}
The radial evolution of these particles is shown in \cref{rtrack}. It demonstrates a similar radial origin for the outflowing components at $3r_{\mathrm{vir}}$ to $4r_{\mathrm{vir}}$. There are, however, also particles that rapidly get kicked to these extreme halo-centric distances despite originally being accreted to well within $1r_{\mathrm{vir}}$ (see the lowest trajectories in the \cref{rtrack}). These particles are most likely ejected by the massive merger that cluster 20 undergoes at late times \citep{kimmig2023} and demonstrates how even processes that happen on relatively long time scales -- such as cluster mergers -- can influence the dynamics on very short timescales and induce large velocities. This is also a channel through which halo-internal processes can directly influence the dynamics in the peripheral environments of these massive clusters and as such is an interesting subject for future work on dynamically very active cluster or supercluster regions.

To compound the information from the purely outflowing component, \cref{T-Z} shows the metallicity--temperature phase space for all particles in the outskirt of cluster 20 between $1.2  r_{\mathrm{vir}}$ and $3.0 r_{\mathrm{vir}}$ (using the final $r_{\mathrm{vir}}$ at $z=0.25$) for different redshifts with blue indicating inflowing and red indicating outflowing particles. There is a clear temperature split between the hot and cold component of outskirt gas. In overall metallicity, the hot outflowing gas is slightly more spread out than the more pristine inflowing gas. (see also: \cref{ztrack}). In the O/Fe-$T$ plane the outflowing gas is distributed almost identically to the inflowing component, providing further evidence that the chemical evolution is largely stopped after infall. Note that in this cluster, which assembles rather late in time, none of the outskirt gas particles from z=0 show significant metal enrichment at z=4. We find that the evolution of the cluster gas elemental abundances starts out at high-redshift being strongly alpha enhanced, matching expectations also for the elements in the stars \citep{kimmig:2025}. As time passes, however, the gas is quickly enriched by SN Ia, strongly reducing [O/Fe] while also increasing [Z], especially within the hot gas phase. This is consistent with cluster 20 lacking a massive protocluster progenitor at z=4. Clusters with high $M_\mathrm{vir}>10^{13}M_\odot$ in contrast, can already host a hot and metal enriched atmosphere, as shown by \citet{remus2023}.

In summary this study of cluster 20 largely confirms the standard picture of cluster accretion, with the gas infalling through dense and relatively cold filaments, enriching on infall and then being heated at the shock surface or below. \Cref{ztrack} clearly shows the chemical development being frozen in after the infalling phase and the subsequent heating is complete, with the final enrichment being dependent on the conditions during infall. We now test these findings across the whole mass sample to verify whether they generalise and are valid across the halo mass range probed in this study or if they might break down at the low-mass end, where the dynamics are more dominated by feedback. 
  
 \begin{figure*}[h!]
	\includegraphics[width=\textwidth]{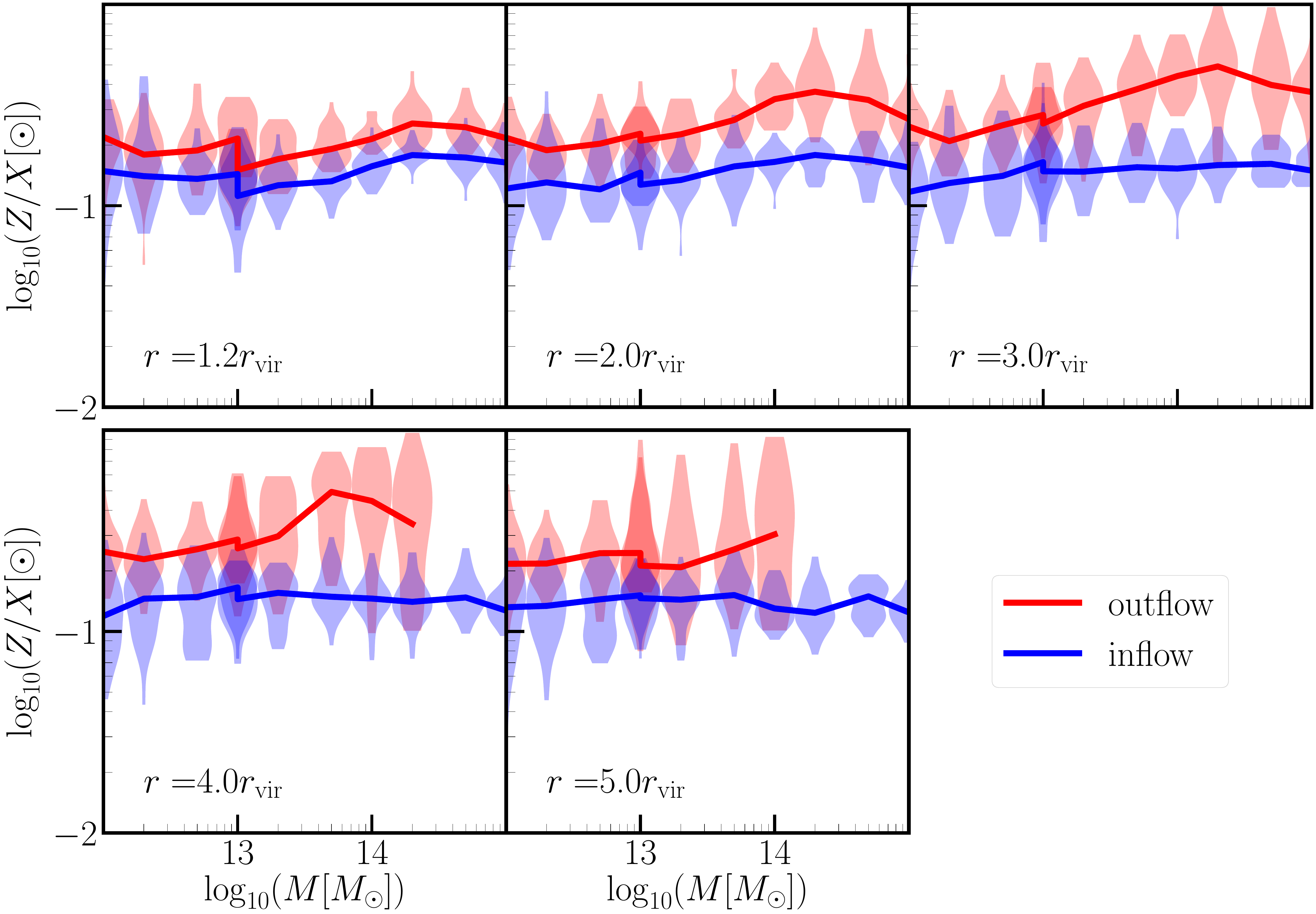}
    \caption{ Metallicity of net (mean) in- and outflow as a function of mass at five radii for the full halo sample. As in \cref{bins3}, the spread within each mass bin is indicated by the vertical distributions. As in \cref{bins3} the cut-off in the high-mass - large radius range does not indicate zero metallicity.}
    \label{bins4}
\end{figure*}

\section{Global Phase properties of the flowing gas} \label{Sec:Phase}
 
 \begin{figure}[!htbp]
  \includegraphics[width=0.999\columnwidth]{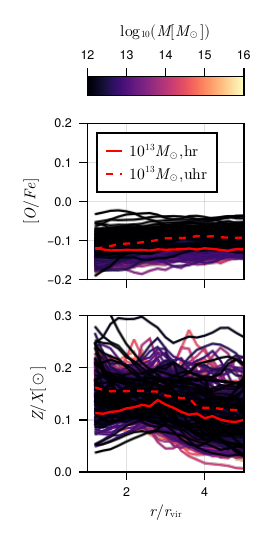}
\caption{Stacked radial O/Fe (top) and total metallicity (bottom) profiles with the same mass colouring as in \cref{entropy}. All haloes from all bins are used.}
\label{ofe}
\end{figure}
We now extend the investigation of the thermal and chemical properties to the whole sample to verify consistency with the findings of the previous section and to analyse the mass scaling behaviour of these properties. 

\subsection{Temperature}
\Cref{bins3} depicts the mean temperature of the gas present in the halo outskirts for the full halo sample split into in- and outflowing gas  (blue and red violins). The red and blue curves indicate the median of the respective mass bin.
While \cref{bins} considered the total particle-based flow, as explained in Sect. 2.6, here only pixels with a \textit{net} in- or outflow are considered for the blue and red curves, respectively. The linearity of the red and blue median curves in \cref{bins3}--in- and outflow respectively--demonstrates a clear mass-scaling behaviour of the gas temperature close to the virial radius. However, there is a gap between the two curves that increases with radius and is more pronounced at the high-mass end. Notably, and inversely to the behaviour of the mass flow shown in \cref{bins}, for the temperature it is the inflowing component that drops away from the mass-scaling relation at $\approx 10^{14} M_{\odot}$ for radii $>1.2 r_{\mathrm{vir}}$, showing a downwards kink. 

The outflow (red curve), in contrast, remains stable and linear out to large radii only decreasing slightly in temperature and showing a radial cut-off at large masses, where the outflows do not reach as far out in terms of scaled distance. This discrepancy in temperature can also be seen spatially in the right panel of \cref{molly3}, exemplary for cluster 20: The regions with a dense net in-flow that dominate the overall mass inflow -- effectively the filaments -- are significantly colder than the regions dominated by the thermally driven outflow. This is a direct manifestation of the shocks discussed in chapter 4. 
 For the clusters of {\it Box2b/hr} and the high-mass group end, the gas falls in at a relatively cold temperature and is then heated when it encounters the pressure supported halo atmosphere. For the less massive haloes in both boxes however, in- and outflow are nearly in temperature equilibrium. This is possibly a manifestation of the mass hierarchy in the two different boxes: Less massive haloes (relative to the maximal halo mass in each box) tend to live in more active environments even when they have not yet been accreted by more massive haloes because there are more equal or higher mass haloes available in their vicinity to heat the surrounding gas. So the apparent drop-off of the inflow temperature at the high-mass end of the boxes should rather be thought of as a relative enhancement of inflow temperature for lower-mass haloes, responsible for the downwards kink in the mass-scaling relation.

\subsection{Metallicity}
 \Cref{bins4} shows the mean metallicity of the gas present in the outskirts of the sample split by flow direction.
 Compared to the clear mass scaling of the temperature curves, the metallicity inflow curves with respect to mass in \cref{bins4} are relatively flat, indicating that the filaments and inflows in general are scale-independent in their chemical evolution. Notable is also a distinct lack of radial evolution, except for the fact that for the inner radial bins, there is a slight increase in metallicity for the massive clusters with decreasing radius. This effect is the strongest for the mass bin centred at $2\times10^{14} M_\odot$. Generally inflows are less metal-rich than outflows, as one would expect. The red outflowing curve has a similar shape to the inflow at small radii. It shows an overall stronger enrichment than the inflowing gas. This is consistent with the picture established in \cref{Sec:Shocks} for the individual cluster study: As gas falls onto massive clusters, it is pre-enriched by the evolving galaxies and groups infalling within the filaments and this evolution stops once the material encounters the shocks surface and is heated. The slight increase in inflowing metallicity with decreasing radius in the high-mass bins also fits into this picture since the gas evolves in metallicity as it falls closer to the central halo. This can also be seen in the tracks in \cref{ztrack} in combination with \cref{rtrack}. 
   \begin{figure*}[!htbp]
  \begin{center}
	\includegraphics[width=\columnwidth]{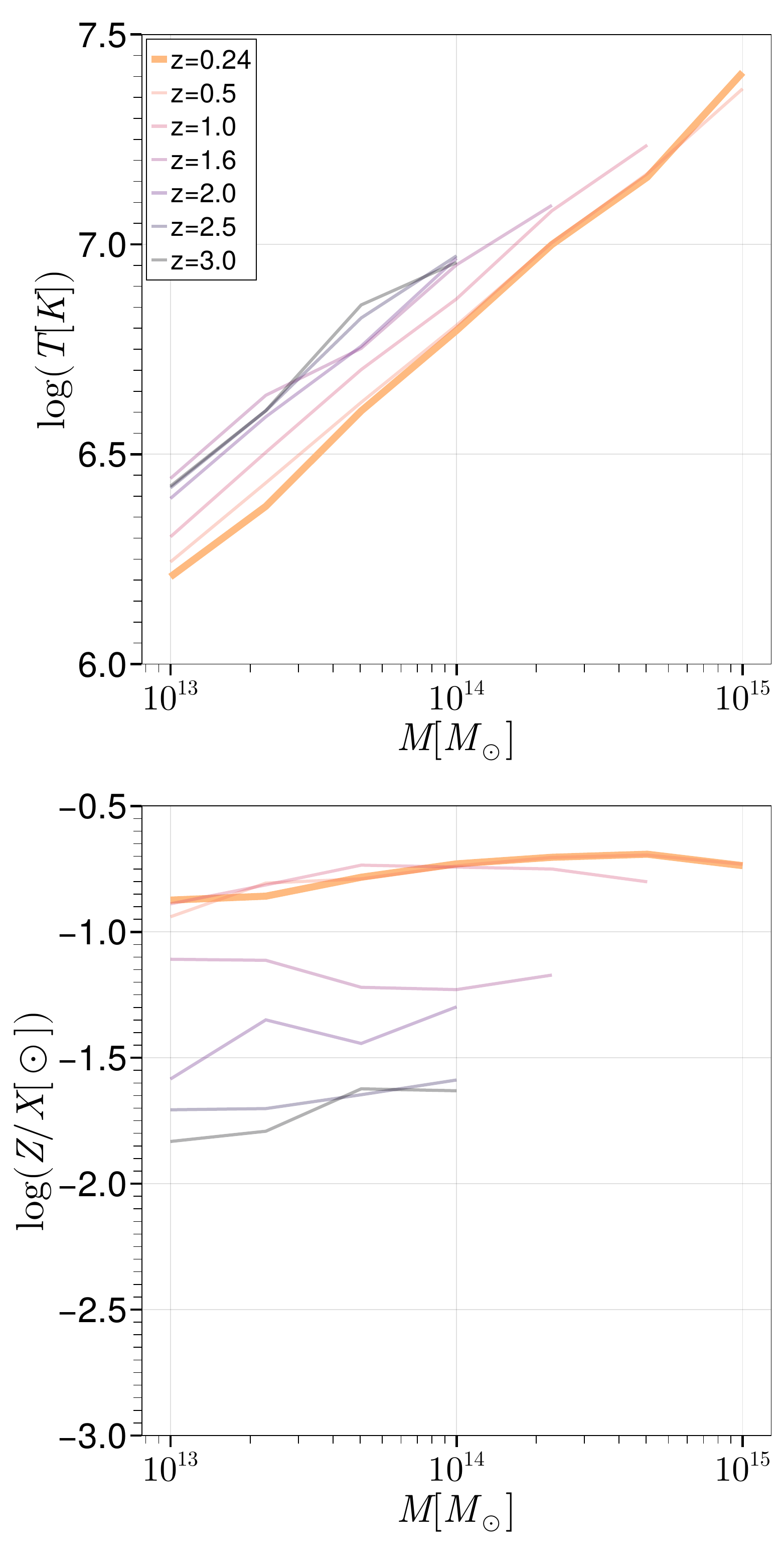}
        \includegraphics[width=\columnwidth]{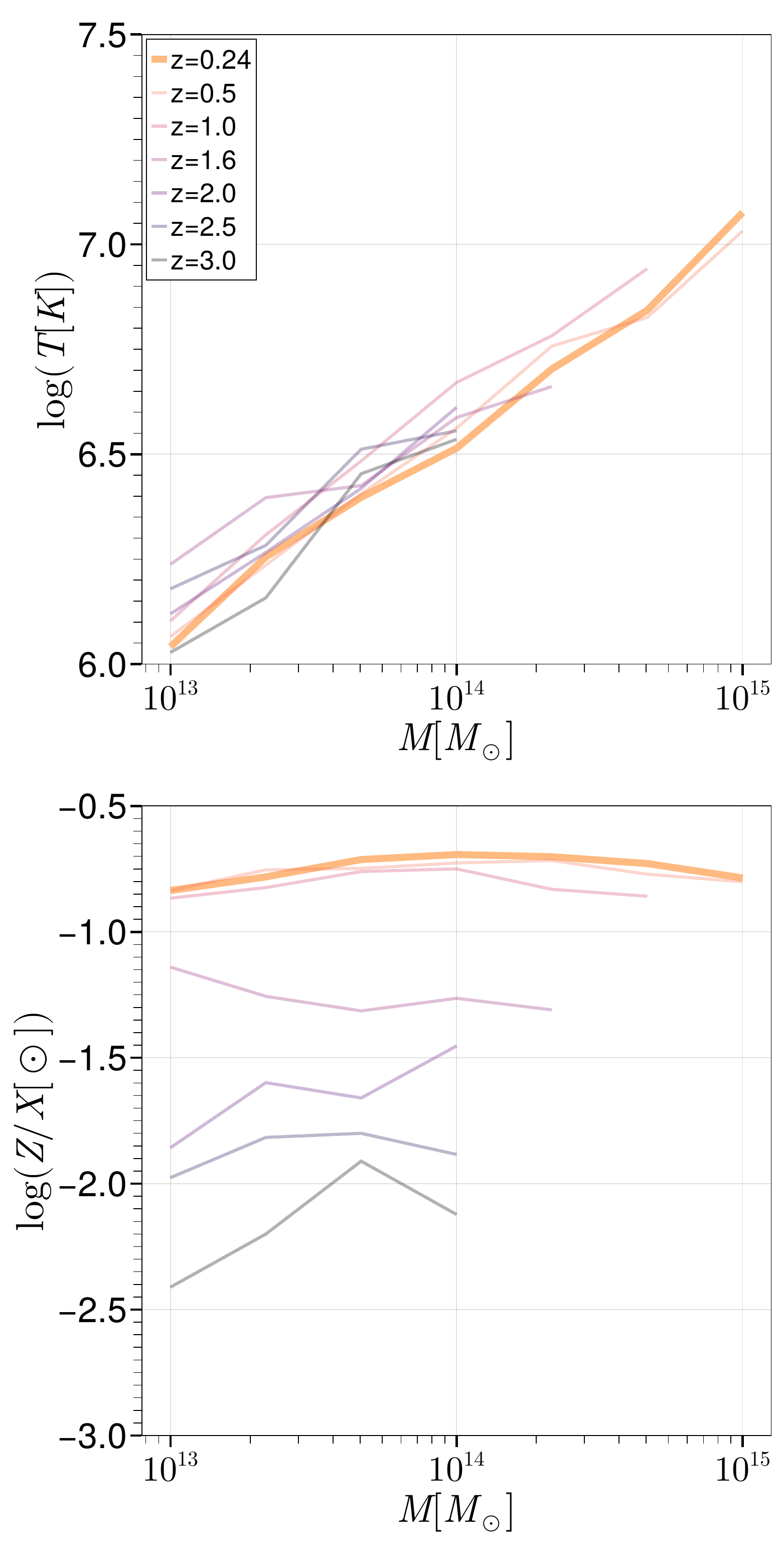}
    \caption{Redshift evolution of the temperature and metallicity at $1.2r_{\mathrm{vir}}$ (left column) and $2.0r_{\mathrm{vir}}$ (right column) as a function of mass. The bin masses are identical to the ones used in the previous section.}
    \label{timevo2}
    \end{center}
\end{figure*}
 
 However, in contrast to the inflowing material, the outflows \textbf{increase} in metallicity going outwards from the central halo for the clusters (mass bins around $10^{14}M_\odot$).
 The increase in metallicity in the outflow for the highest mass bin is possibly a contamination effect of massive neighbours of these clusters contributing to the outflow with their own accretion (dominating the mass averages due to the inflows being more dense overall). 
 The metallicity curves in \cref{bins4} show a stronger offset relative to the vertical scatter within the bins between outflowing and infalling material when compared to the temperature in \cref{bins3}. There additionally is, similarly to the temperature, a noticeable discontinuity in metallicity between the two simulations at the mass bin where they overlap at $10^{13}M_\odot$. Present in both components at small radii, this jump vanishes further out. While {\it Box2b/hr} and {\it Box4/uhr} differ in their global star formation and thus in their enrichment properties due to their different resolutions, the "floor" of enrichment is the same for both boxes in the relatively pristine and low density regime of the furthest halo outskirts. At around $0.1 Z_\odot$, most enriched gas particles have only been impacted by one to a few supernovae during their evolution independently of the resolution, explaining the agreement between the boxes at large radii. In the outflowing component there is a stronger residual resolution gap, consistent with the picture of the outflowing gas at large radii being the gas that is enriched and pushed out during the early formation of the central galaxy, where the resolution would indeed impact the efficiency of star formation and hence enrichment. 
The top panel of \cref{ofe} shows the radial oxygen-to-iron ratio profiles for all haloes in the sample. The profiles are consistently flat across the entire mass range and are in agreement with previous numerical \citep{biffi2017,vogelsberger2018,angelinelli2022} and observational \citep{sakuma2011} studies on the inner regions of the ICM. There is however an apparent overall mass trend with higher mass haloes having lower oxygen to iron ratios at each radius. This trend is likely caused by the difference in resolution between {\it Box4/uhr} and {\it Box2b/hr} as the red median lines calculated with the $10^{13}M_\odot$ bin in both boxes indicate. Previous works have examined differences in the cosmological star formation rates and their histories between the resolution levels (e.g. Dolag et al. 2025, submitted) and these differences will plausibly also impact the chemical composition of the gas, especially the O/Fe ratio, which is directly sensitive to the timescales of star formation in the cosmological volume. The overall metallicity in the bottom panel of \cref{ofe} demonstrates a global trend of similar consistency across the mass range. There is a slight negative metallicity gradient for all haloes evening out eventually as the chemical influence of the central halo vanishes at large radii. ($4r_{\mathrm{vir}}<r<5r_{\mathrm{vir}}$). Overall these findings (flat O/Fe ratio, slightly negative radial metallicity gradient and increased temperature for the outflowing component) are consistent with the early enrichment and subsequently frozen-in chemical evolution in the outskirt gas found for cluster 20. To complete the picture from the individual cluster study, it is also necessary to consider the temporal evolution of the scaling relations from \cref{bins3} and \cref{bins4}, which we discuss in the following section.
 \subsection{Redshift evolution of temperature and metallicity}
\Cref{timevo2} shows the evolution of the halo mass--gas temperature (upper panels) and halo mass--gas metallicity (lower panels) relation close to the virial surface and further in the outskirts. The temperature close to the virial surface decreases consistently across the mass range with time, keeping the mass scaling stable. This is still consistent with haloes and in consequence halo outskirts heating dynamically as they gain mass through mergers and accretion: As we consider equal-mass bins in time as opposed to tracing individual haloes, growing haloes will move from one bin to another in time and in consequence up the temperature mass relationship. The cooling trend in the temperature-mass relation rather reflect an overall cooling of the immediate surroundings of these objects as they evolve and accrete mass. In contrast, there is no clear time evolution of the mass-temperature scaling relationship at larger radii, which could plausibly be erased by the many factors that the temperature of the larger-scale environments of these objects depends on. 

The metallicity evolution (lower panels) shows the picture of early enrichment and subsequent stagnation of the chemical evolution in the outskirt gas to be consistent across the mass range: At both sampled radii the normalisation of the metallicity-mass relation only significantly changes at $z>1.0$. At later times the previously demonstrated flat mass-metallicity curve stabilises, showing little variation due to the enrichment in the outskirts having predominantly taken place at early times.   

\section{Conclusions}
We have used the Magneticum suite of simulations to examine large-scale matter flows in the outskirts of haloes spanning a large range of masses and to study the mass scaling behaviour of these flows. This allowed us to determine how self-similar the accretion and outflow processes are across the entire range of masses. The main findings of this work are:
\begin{itemize}
  \item While the individual spatial flow configurations are complex and vary strongly from halo to halo, the average inflows onto haloes in the simulation follow a mass scaling relation that is consistent with the semi-analytical model for a hierarchically accreting overdensity and that is stable in time. This is important because it implies that inputs for mass-accretion models that do not emphasise the exact configuration of the flow fields (e.g., the Bathtub model for the balance of mass accretion and star formation \citep{dekel2013}) can be kept relatively simple and consistent across scales.
  \item Considering the azimuthal symmetry of the inflow and outflow fields separately, inflows demonstrate a significantly less isotropic distribution when compared to outflows. This is consistent at all scales, though possibly the isotropy of outflows at small scales is overstated due to the feedback model used in the simulations not implementing kinetic feedback which might introduce additional anisotropy at these scales. These findings confirm the expectations of inflows being largely filamentary from coupling to the large scale cosmic web, where the tidal configuration aids the pre-collapse of these elongated structures. The consistency of this inflow anisotropy confirms this to also be true for smaller haloes, being fed by smaller sub-filaments given by the fractality of the cosmic web itself.
  \item The thermal effects break the self-similar mass scaling of the flow fields for the outflowing component. The stronger impact of AGN feedback on galaxy to group size haloes can be seen as a characteristic kink in the mass scaling relation as low-mass haloes show relatively stronger outflows at larger radii ($r>3r_{\mathrm{vir}}$), extending previously found trends in simulations for clusters and groups in the mass scaling of the closure radius across an even larger mass range \citep{angelinelli2023}.
  \item In the Magneticum simulations galaxy and group-mass objects have flat entropy profiles in the outskirts to larger radii as the gas is mainly heated by the feedback processes. In contrast, cluster-mass objects show entropy peaks that reach out to $\approx 2-3 r_{\mathrm{vir}}$, consistent with expectations for MA-shocks \citep{zhang2020} in the outskirts of clusters that have undergone a massive merger recently. We report a decrease of entropy peak radius with mass for the objects that do have a significant entropy peak, consistent with previous work. 
  \item The metals present in the outskirts of the example cluster we studied in detail are predominantly formed during gas infall and before the gas encounters the shock surface and thermalizes its kinetic energy, confirming the early enrichment scenario \citep{biffi2018}. Both the total metallicity and the O/Fe mass ratio as a proxy to enrichment history effectively stop evolving once the gas joins the hot expanding atmosphere in the cluster outskirts.
  \item The temperature in the cluster outskirts shows a consistent mass scaling behaviour of mass increasing with temperature with the outflows being consistently hotter than the cold inflowing gas due to shock heating and AGN feedback.
  \item The degree of enrichment of galaxy cluster and galaxy outskirts is very similar with no clear mass scaling behaviour. Overall the outflowing component demonstrates a higher degree of enrichment compared to the inflow, and the radial profiles are consistent with the ones found in previous work. We however report a peculiar increase of the metallicity of outflowing gas for cluster haloes in the intermediate to large radial range ($2-4r_{\mathrm{vir}}$).
\end{itemize}

\begin{acknowledgements}
BAS acknowledges support by the grant agreements ANR-21-CE31-0019 / 490702358 from the French Agence Nationale de la Recherche / \emph{Deut\-sche For\-schungs\-ge\-mein\-schaft, DFG\/} through the LOCALIZATION project. LMV acknowledges support by the German Academic Scholarship Foundation (Studienstiftung des deutschen Volkes) and the Marianne-Plehn-Program of the Elite Network of Bavaria. LCK acknowledges support by the DFG project nr. 516355818. KD acknowledges support by the COMPLEX project from the European Research Council (ERC) under the European Union’s Horizon 2020 research and innovation program grant agreement ERC-2019-AdG 882679.
\end{acknowledgements}
%
\bibliographystyle{aa} 
\bibliography{bib.bib} 
%

\end{document}